\documentclass[aps,prb,showpacs,floatfix,twocolumn,superscriptaddress]{revtex4-2}

\usepackage{graphicx}
\usepackage{dcolumn}
\usepackage{booktabs}
\usepackage{siunitx}
\usepackage[english]{babel}
\usepackage{mathtools}  % 包含amsmath
\usepackage{amssymb}
\usepackage{bm}
\usepackage{cancel}
\usepackage{nicefrac}

\usepackage{braket} %  <>

\usepackage{tikz}

\usepackage{tablefootnote}
\usepackage[%
pdfauthor={Benedikt Placke},
pdfstartview=FitH,
breaklinks=true,
bookmarks=true,
colorlinks=true,
anchorcolor=black,
citecolor=red,
filecolor=black,
menucolor=black,
urlcolor=blue,
linkcolor=red,
]{hyperref}
\usepackage{upgreek}
\begin{document}
	
	% \title{Phase-dependent nonequilibrium scaling across BKT transitions in the $J_1$--$J_2$  Potts Model}

\title{Logarithmic scaling correction in quench dynamics of the 
\texorpdfstring{$J_1$--$J_2$}{J1--J2} Potts model}

	\author{Kun Li}
	\affiliation{College of Physics, Taiyuan University of Technology, Shanxi 030024, China}
	
	\author{Wanzhou Zhang}
    \email{zhangwanzhou@tyut.edu.cn}
	\affiliation{College of Physics, Taiyuan University of Technology, Shanxi 030024, China}
	% \author{Chengxiang Ding}
	% \email{dingcx@ahut.edu.cn}
	% \affiliation{School of Microelectronics $\&$ Data Science, Anhui University of Technology, Maanshan 243002, China}
	
	\date{\today}
	
	\begin{abstract}
    %常见淬火动力学对应的KZ机制对应缺陷密度随着淬火速率满足幂律衰减。具有拓扑相变的XY模型的KZ机制幂律关系存在对数修正。然而对于其他离散自旋的且改变温度时发生两次拓扑相变是否存在对数修正未知。本文以J1-J2 Potts模型为例，给出了其平衡态相图，并根据其三个相（顺磁相，准长程序相，长程序相）以及零温基态的温度范围设置了四个不同温度区间的淬火协议。结果表明，
    In conventional quench dynamics governed by the Kibble–Zurek mechanism (KZM), the defect density generally decays as a pure power law of the quench rate. However, the KZM scaling of the two-dimensional (2D) XY model with topological phase transitions features prominent logarithmic corrections. Nevertheless, it remains unclear whether such logarithmic scaling corrections emerge in discrete-spin systems that host two successive topological phase transitions under thermal quenches.
    This work investigates the \(J_1\)–\(J_2\) antiferromagnetic Potts model and constructs its equilibrium phase diagram. Based on the temperature ranges of the paramagnetic phase (PM), quasi-long-range ordered (QLRO) phase, long-range ordered (LRO) phase, and zero-temperature ground state, we design four quench protocols with distinct temperature intervals.
    Our results demonstrate that
    quenches terminating in the QLRO phase exhibit logarithmically corrected KZM scaling of the excess energy density, consistent with the dynamical universality class of the 2D XY model.
     In contrast, quenches ending in the LRO phase, including both finite-temperature and zero-temperature protocols, follow conventional power-law scaling.  Our results clearly uncover the characteristic scaling corrections of the \(J_1\)–\(J_2\) Potts model and offer theoretical guidance for future experimental investigations of KZM via photonic simulation platforms.
    
    % We investigate the nonequilibrium quench dynamics of the two-dimensional three-state $J_{1}$--$J_{2}$ antiferromagnetic Potts model using Monte Carlo simulations. Equilibrium thermodynamic analysis identifies two successive Berezinskii-Kosterlitz-Thouless (BKT) transitions, separating the paramagnetic (PM), quasi-long-range ordered (QLRO), and long-range ordered (LRO) phases. Based on the resulting phase diagram, we systematically study four representative finite-rate quench protocols. We find that the nonequilibrium scaling is determined primarily by the destination phase of the quench. Quenches terminating in the QLRO phase exhibit logarithmically corrected Kibble-Zurek scaling of the excess energy density, consistent with the dynamical universality class of the two-dimensional XY model. In contrast, quenches ending in the LRO phase, including both finite-temperature and zero-temperature protocols, follow conventional power-law scaling without logarithmic corrections. The nonequilibrium squared staggered magnetization exhibits distinct scaling behaviors under different quench protocols. Our results reveal a clear phase-dependent crossover between logarithmically corrected and conventional Kibble-Zurek scaling in the $J_{1}$--$J_{2}$ antiferromagnetic Potts model.
	\end{abstract}

	\maketitle
	
	\section{Introduction}
The Kibble–Zurek mechanism (KZM) predicts that finite-rate quenches across a continuous phase transition generate topological defects with a steady power-law scaling between defect density and quench rate, where the universal scaling exponent is uniquely determined by the equilibrium critical behavior of the system~\cite{TWBKibble_1976,Zurek:1985qw}. For a linear quench protocol, the control parameter departs from the critical point \(T_c\) as \(T(t)-T_{c}=Rt\), where \(R=1/\tau_Q\) denotes the quench rate and \(\tau_Q\) is the quench time. Within the standard KZM framework, the resulting excess topological defect density \(\delta n\) obeys the scaling relation
\begin{equation}
\delta n \propto  R ^{(D-d) \nu/(1+ \nu z)},
\end{equation}
where $D$ is the spatial dimension of the system, $d$ is the dimension of topological defects, \(\nu\) is the correlation-length critical exponent, and $z$ is the dynamical critical exponent~\cite{doi:10.1142/S0217751X1430018X}. A core consequence of renormalization-group theory is that systems in the same universality class exhibit identical critical exponents and dynamical scaling properties, regardless of their microscopic differences~\cite{RevModPhys.46.597,RevModPhys.70.653,RevModPhys.49.435,Sachdev1998}. This strong universality renders the KZM a fundamental paradigm for characterizing nonequilibrium critical dynamics, which has been broadly verified in both theoretical simulations and experimental realizations. The validity of KZM scaling has been numerically confirmed in classical statistical models~\cite{PhysRevLett.132.241601,PhysRevB.90.134108,Jelić_2011,PhysRevE.99.022113,t1f5-prsf}, closed quantum many-body systems~\cite{PhysRevB.104.035423,PhysRevA.101.023610,PhysRevB.106.184301,PhysRevB.92.035117,PhysRevB.102.134302,PhysRevLett.116.225701,PhysRevLett.118.065701,PhysRevB.110.045140,PhysRevA.108.023312,PhysRevB.104.014406}, and dissipative open quantum systems~\cite{PhysRevB.111.184319,PhysRevB.111.155152}. Experimentally, state-of-the-art quantum simulators, including ultracold atomic gases and Rydberg atom arrays, have precisely measured the universal KZM power-law scaling of defect formation~\cite{PhysRevLett.116.155301,PhysRevLett.117.275701,PhysRevA.89.022337,PhysRevLett.105.075701}. Beyond microscopic quantum and superfluid platforms, the KZM universality further holds for macroscopic classical dissipative systems. Representative  experimental verifications include Rayleigh–Bénard thermal convection, which confirms KZM power-law defect scaling in multi-mode hydrodynamic structures~\cite{PhysRevE.74.047101}, and anisotropic structural phase transitions on Si(001) surfaces~\cite{rmc4-xqb3}. %Both experiments consolidate the generality of the KZM paradigm across vastly different length scales and dissipation regimes.

\begin{figure}[t]
    	\includegraphics[width=0.45\textwidth]{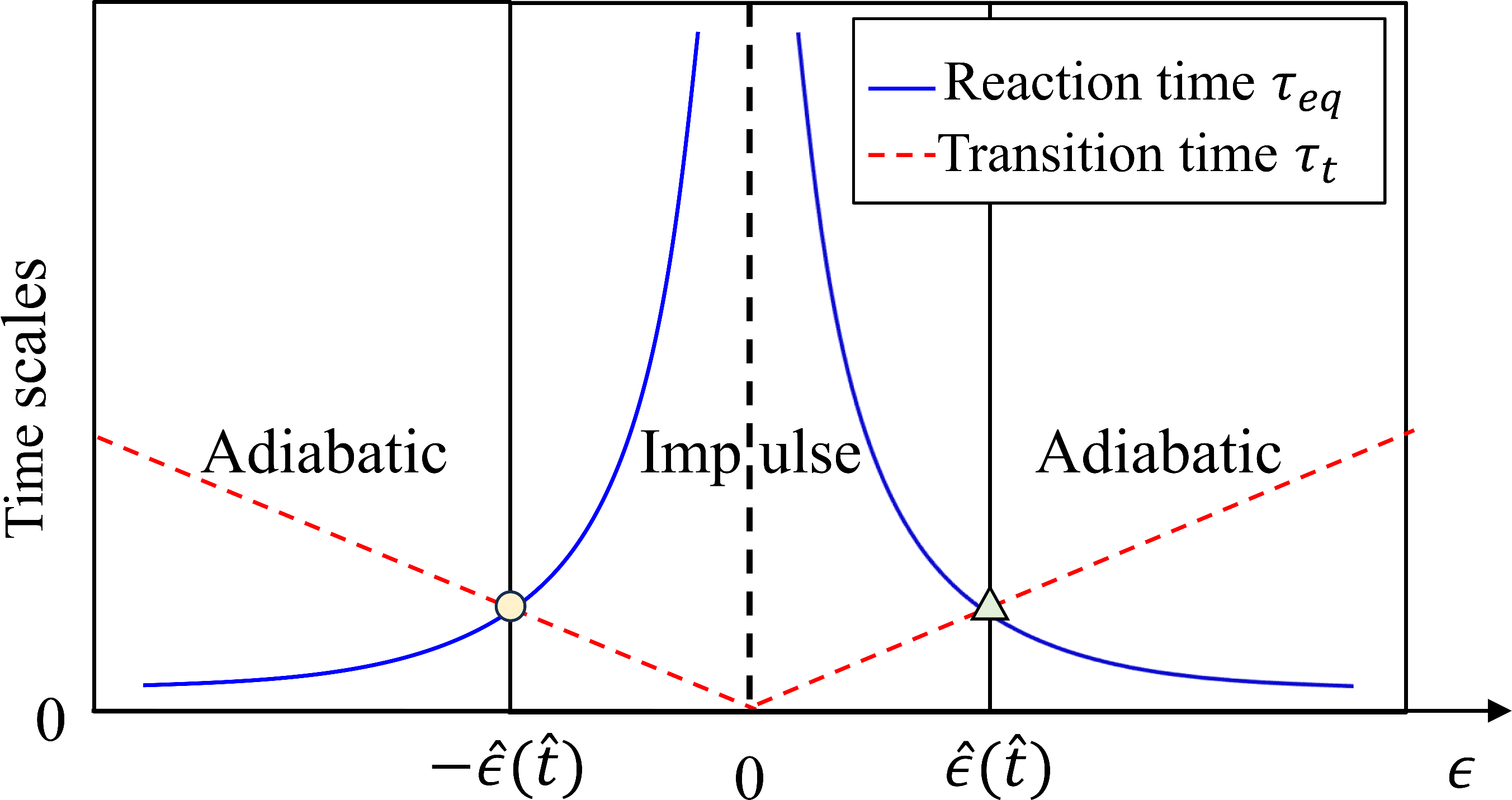}
    	\caption{Schematic illustration of the KZM. The reaction time $\tau_{eq}$ and the transition time $\tau_{t}$ are plotted as functions of the dimensionless distance $t$ from the critical point. The intersection points of these two curves, marked by light orange circles and light green triangles at $-\hat{t}$ and $\hat{t}$ respectively, define the crossover between the adiabatic and impulsive regimes of the time evolution.}
    	\label{fig:KZ}
    \end{figure}
The universal power-law scaling of the standard KZM holds only under a strict set of assumptions: the transition must be continuous, the system resides in the thermodynamic limit, no explicit symmetry-breaking external fields are applied, and the quench rate lies within the regime where the adiabatic-impulse approximation remains valid~\cite{RevModPhys.83.863}  in Fig.~\ref{fig:KZ}. Breaking any of these conditions breaks the simple power-law relation for topological defect densities. A large body of recent theoretical and experimental work has explored non-KZM dynamical responses induced by these violations, uncovering rich exotic nonequilibrium critical behaviors. Four major sources of scaling breakdown are widely studied:
(i) ultrafast quenches that invalidate the adiabatic-impulse picture~\cite{PhysRevB.108.214307,PhysRevLett.130.060402};
(ii) initial states far from thermal equilibrium or the critical point~\cite{PhysRevB.93.024103};
(iii) spatially nonuniform driving protocols across the sample~\cite{PhysRevLett.125.260603,PhysRevLett.105.075701,PhysRevA.103.013310,PhysRevA.108.023315};
(iv) dissipative noise and environmental coupling in open quantum systems~\cite{PhysRevLett.124.230602,PhysRevB.95.224303}.
Such deviations are not limited to minor shifts of the effective scaling exponent; they can qualitatively reshape the entire functional form of the scaling law. Common sub-leading corrections to ideal KZM power laws include exponential prefactors~\cite{clvs-yk7v,PhysRevB.111.155152,bacsi2023kibble}, environment-induced correction terms~\cite{5xrs-z9ls}, and, logarithmic scaling modifications.

One prominent correction to the power-law scaling of the KZM is the logarithmic correction, which introduces an additional logarithmic factor that alters the observable’s dependence on the quench rate. For quenches across the Berezinskii–Kosterlitz–Thouless (BKT) transition of the two-dimensional (2D) XY model, vortex densities carry such logarithmic corrections as a function of the quench rate~\cite{Jelić_2011}. Similarly, the equilibrium correlation length of a disordered quantum transverse-field Ising chain shows weak logarithmic scaling with the quench rate~\cite{Dziarmaga2014quench,PhysRevB.74.064416,PhysRevB.76.144427,PhysRevB.79.094421}, and vortex dynamics in Raman spin-orbit-coupled Bose–Einstein condensates under thermal quenches also feature logarithmic sub-leading terms~\cite{qlx5-dqr3}. Logarithmic modifications to vortex density scaling have been reported in finite-rate quenches of 2D superfluids~\cite{PhysRevLett.110.165303}. The origin of these logarithmic scaling corrections are related with the quasi-long-range ordered (QLRO) low-temperature phase that emerges below the BKT transition~\cite{Jelić_2011}. %\textcolor{blue}

Within this QLRO phase, the binding energy of a vortex–antivortex pair separated by some distance scales logarithmically with the pair separation. This long-range interaction generates logarithmic corrections to the equilibrium correlation lengths~\cite{RevModPhys.89.040501,JMKosterlitz_1972}, which further propagate to the scaling of order parameters including vortex density and magnetization. For classical spin systems, however, existing work on logarithmic KZM corrections concentrates almost exclusively on systems with continuous spin symmetries, such as the 2D XY model,  which host only a single BKT transition. \textit{Whether analogous logarithmic corrections on KZM can arise in lattice models with discrete spin degrees of freedom remains an open question.}

% As exemplified by the $J_1$–-$J_2$ Potts model~\cite{RevModPhys.54.235}, all spin degrees of freedom therein take discrete finite values. Prior works~\cite{PhysRevLett.46.1458,PhysRevB.26.6326,etde_20825529} concentrated exclusively on individual equilibrium quantities without a systematic survey of the full equilibrium phase diagram, and no evidence for logarithmic corrections has been observed in existing literature.
% To address this gap, we investigate the finite-rate quench dynamics of the two-dimensional $J_{1}$–-$J_{2}$ three-state Potts model using Monte Carlo (MC) simulations~\cite{10.1063/1.1699114}. We first establish the equilibrium phase diagram for $J_{1}=-2$ and $J_{2}=1$, demonstrating the existence of three distinct thermodynamic phases—a paramagnetic (PM) phase, a QLRO phase, and a long-range ordered (LRO) phase—separated by two BKT transitions. To probe the nonequilibrium dynamics associated with these phases, we design four quench protocols: PM $\to$ QLRO, PM $\to$ LRO, PM $\to$ $T=0$, and QLRO $\to$ LRO. 

As exemplified by the $J_1$–$J_2$ Potts model~\cite{RevModPhys.54.235}, spin degrees of freedom in such systems take discrete, finite values. Previous studies~\cite{PhysRevLett.46.1458,PhysRevB.26.6326,etde_20825529} have focused solely on individual equilibrium observables rather than systematic characterization of the full phase diagram, leaving logarithmic scaling corrections largely unexplored in  $J_1$–$J_2$ Potts systems. To address this open issue, we investigate the finite-rate quench dynamics of the 2D $J_1$–$J_2$ three-state Potts model via Monte Carlo (MC) simulations~\cite{10.1063/1.1699114}. We first construct the equilibrium phase diagram at $J_1=-2$ and $J_2=1$, identifying three distinct thermodynamic phases: a paramagnetic (PM) phase, a QLRO phase, and a long-range ordered (LRO) phase, which are separated by two successive BKT transitions. To systematically explore the nonequilibrium scaling behaviors across different phase regimes, we implement four representative quench protocols: PM $\to$ QLRO, PM $\to$ LRO, PM $\to$ $T=0$, and QLRO $\to$ LRO.

% \textcolor{blue}{For quenches terminating in the QLRO phase, we observe logarithmically corrected scaling laws whose scaling forms and characteristic exponents are consistent with those reported for the two-dimensional XY model. In contrast, quenches terminating in the LRO phase exhibit conventional power-law scaling without logarithmic corrections, with scaling exponents consistent with those expected for three-state Potts ordering dynamics}.

% \textcolor{blue}{By systematically comparing the scaling behavior across different quench trajectories, we demonstrate that distinct thermodynamic phases are governed by qualitatively different nonequilibrium scaling laws. These results reveal a direct connection between equilibrium phase structure and nonequilibrium universality, and further show that the excess energy density $\delta E(t)$ and the squared staggered magnetization $\langle M_s^2(t) \rangle$ can exhibit markedly different scaling behaviors depending on the quench protocol and the phase reached during the evolution.}

From observations of the excess energy density \(\delta E(t)\), quenches terminating in the QLRO phase obey scaling laws with logarithmic corrections, whose functional forms and characteristic exponents agree with published results for the 2D XY model. In contrast, quenches that end in the LRO phase exhibit pure standard power-law scaling free of logarithmic corrections. We additionally examine the squared staggered magnetization $\langle M_s^2(t)\rangle$.% whose scaling behavior differs markedly from that of \(\delta E(t)\).

This paper is organized as follows. 
Section~\ref{sec:KZ MECHANISM AND BKT SCALING} reviews the KZM and BKT scaling theory. Section~\ref{sec:MODEL AND  METHOD} details the model and observables. Section~\ref{sec:results} presents our numerical results for the equilibrium phase diagram and nonequilibrium scaling for the different quenching protocol process. Conclusions are summarized in Sec.~\ref{sec:CONCLUSION}.    
\section{ KZM SCALING THEORY}
\label{sec:KZ MECHANISM AND BKT SCALING}
\subsection{KZM Scaling Theory Without Logarithmic Correction}\label{sec:KZM}  
According to the KZM, the dynamical evolution of a system subjected to a finite-rate quench can be divided into three distinct regimes: two adiabatic regimes and one impulse regime, as illustrated in Fig.~\ref{fig:KZ}.

During a finite-rate quench, particular attention is paid to the moment when {the reaction time $\tau_{eq}$} equals the drive transition {time scale $\tau_{t}$}, known as the freeze-out time $\hat{t}$, as given below,
\begin{equation}\label{eq:2:HATT}
\tau_{eq}(|\hat{t}|) =\tau_{t}(|\hat{t}|).   
\end{equation}
In this expression, $\tau_{eq}$ refers to the characteristic time for spins to rearrange and reach equilibrium. The quantity $\tau_{t}$
%(the inverse transition rate,|$\epsilon/\hat{\epsilon}$|) 
is the characteristic time scale of the external driving, representing the time required for the control parameter to reach the critical point at the current transition rate.
Topological defects effectively emerge around $t=-\hat{t}$, and accumulate during the impulse regime. Around $t=\hat{t}$, the defect density reaches its maximum value, which is governed by the frozen correlation length $\hat{\xi}$. The excess defect density is defined as $ \delta n \equiv n(t)-n^{\mathrm{eq}}[T(t)]$, and the corresponding scaling relation is given by:  
\begin{equation}\label{eq:delta ndt_zhengwen}
    \delta n \sim \hat{\xi}[T(\hat{t})]^{-(D-d)} \sim \tau_{Q}^{-(D-d) \nu / (1+ \nu z)},
\end{equation}
and $\hat{t}$ scaling form,
 \begin{equation}\label{eq:thatscaling}
  \hat{t} \sim \tau_{Q}^{\nu z/(1+\nu z)}.
\end{equation}
The derivation of Eqs.~(\ref{eq:delta ndt_zhengwen}) and (\ref{eq:thatscaling}) is summarized in Ref~\cite{Dziarmaga2010}.

The conventional KZM predicts a pure power-law scaling of the defect density with the quench time. For systems undergoing a BKT transition, however, this scaling can be modified by logarithmic corrections. A prototypical example is provided by the 2D XY model, for which the corresponding nonequilibrium scaling relations have been derived analytically and verified numerically~\cite{Jelić_2011}. For later comparison with our results, we briefly summarize these scaling forms below,
\begin{align}
& \hat{t} \simeq  \tau_{Q} \left[ \frac{b_{\tau} z}{\ln(\tau_Q /\tau_0)} \right]^{1/\nu},  \
& \delta n \simeq \left[\frac{\tau_Q}{\ln\left( \tau_Q / \tau_{0} \right)} \right]^{-1},
\label{eq:rho_qlro_8}
\end{align}

where \(b_{\tau}\) and \(\tau_{0}\) are non-universal microscopic time scales. In large $\tau_{Q}$ limit, \(\tau_Q \gg \tau_0\), the logarithmic term \(\ln(\tau_Q/\tau_0)\) varies slowly and acts as a weakly varying prefactor, such that \(\hat{t}\) is approximately proportional to \(\tau_Q\), namely
\begin{equation}\label{eq:hatt}
\hat{t} \simeq \tau_{Q}.
\end{equation}
%While the logarithmic factor only induces a mild correction to the scaling of \(\hat{t}\) and can be safely neglected in practical fitting, it constitutes a dominant slow-varying correction term for the defect density \(\delta n\). 
Eq.~\eqref{eq:hatt} indicates that the freeze-out time scales linearly with the quench time, whereas Eq.~\eqref{eq:rho_qlro_8} demonstrates that the defect density carries a  logarithmic correction on the standard KZM power law. 
% \textcolor{red}{ Eq.~\eqref{eq:rho_qlro_8} shows the defect density follows a new logarithmic scaling formula, rather than  the conventional KZM power law.}
Throughout this paper, \(\simeq\) denotes full asymptotic approximations that retain constant prefactors and logarithmic slow corrections. The derivation of these scaling relations is  outlined in Ref~\cite{Jelić_2011}.

%In the following sections, Eqs.~\eqref{eq:hatt} and \eqref{eq:rho_qlro_8} will serve as the theoretical benchmarks for identifying logarithmically corrected scaling behavior in the quench dynamics of the $J_{1}$–-$J_{2}$ three-state Potts model.    
\section{MODEL AND  OBSERVABLES}\label{sec:MODEL AND  METHOD} 
\subsection{Model}

The Hamiltonian for the $q=3$ three-state Potts model defined on a 2D square lattice with competing nearest-neighbor (NN) and diagonal next-nearest-neighbor (NNN) interactions reads
\begin{equation}\label{H}
	H=-J_1\sum_{\langle i,j\rangle} \delta_{\sigma_i \sigma_j} -J_2\sum_{\langle\langle i,j\rangle\rangle} \delta_{\sigma_i \sigma_j},
\end{equation}
where each lattice spin $\sigma_i$ takes integer values $\sigma_i \in \{1,2,3\}$. The bracket notation $\langle i,j\rangle$ denotes NN spin pairs, while $\langle\langle i,j\rangle\rangle$ stands for diagonal NNN spin pairs, and $\delta_{\sigma_i \sigma_j}$ is the Kronecker delta function. Throughout this work, we adopt competing couplings $J_1=-2$ (antiferromagnetic NN) and $J_2=1$ (ferromagnetic NNN).

\begin{figure}[t]
	\includegraphics[width=0.45\textwidth]{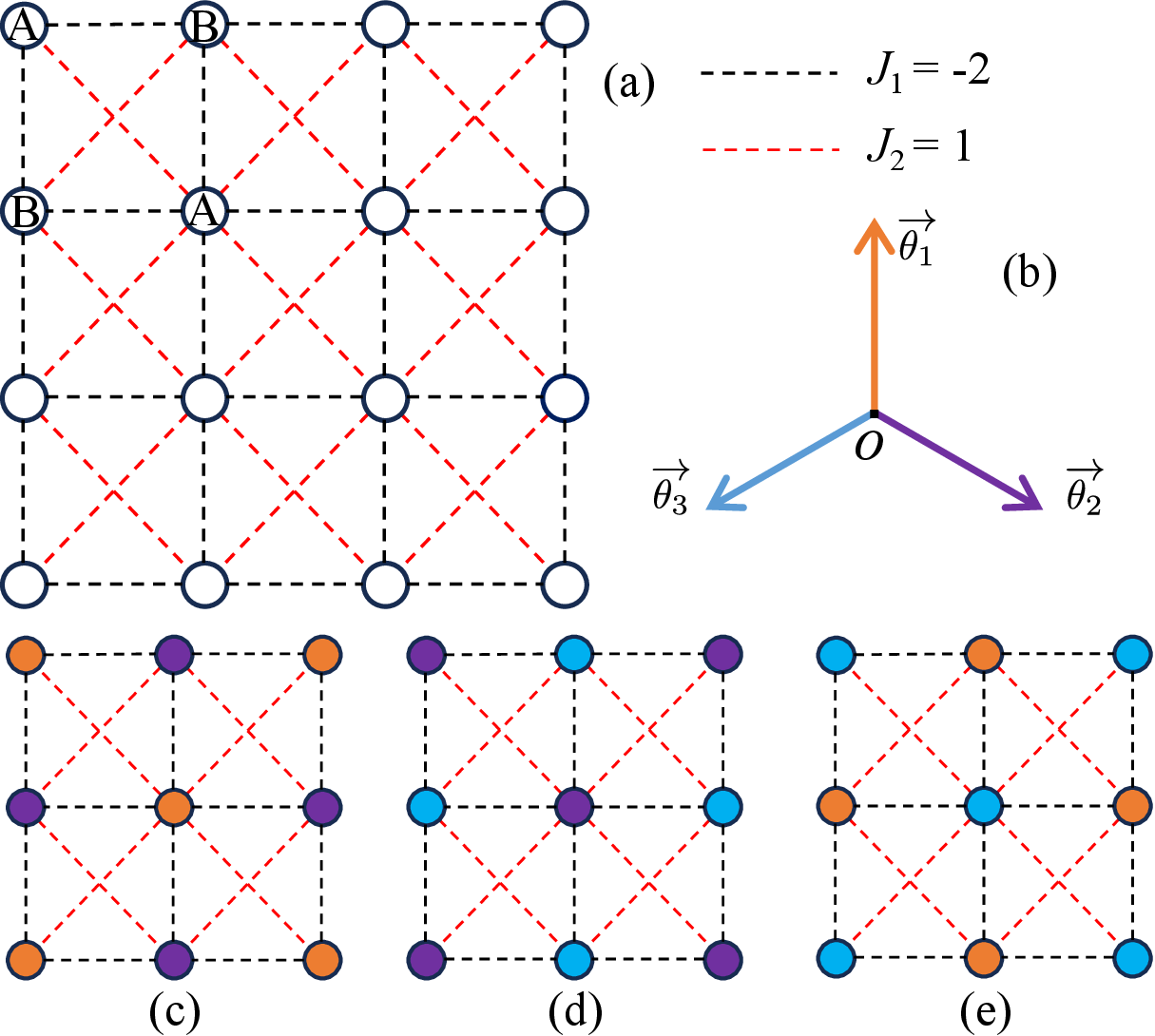}
	\caption{Lattice and spin basis setup for the \(J_1\)–\(J_2\) three-state Potts model studied in this work. (a) Two types of pairwise spin interactions on the square lattice: gray dashed bonds are NN couplings \(J_1=-2\), while red bonds correspond to NNN interactions \(J_2=1\). Labels A and B mark the two sublattices of the bipartite square lattice. (b) Three unit vectors \(\vec{\theta}_{1,2,3}\) forming the basis of the three-state Potts spin space. (c)–(e) Three of the six degenerate ground-state configurations of the model. The remaining three are obtained by exchanging two Potts spin states.
}
	\label{fig:lattice_and_GS}
\end{figure}

We map each discrete Potts spin state to an angular variable on the unit circle via the relation
\begin{equation}
	\theta_i = (\sigma_i-1)\,\frac{2\pi}{3},\quad \sigma_i \in \{1,2,3\}.
\end{equation}
Here, $\sigma_i$ labels the spin state at lattice site $i$, corresponding to the three basis vectors illustrated in Fig.~\ref{fig:lattice_and_GS}(b). The angle $\theta_i$ yields Cartesian spin projections $\cos\theta_i$ and $\sin\theta_i$, which we use to compute key observables such as the staggered magnetization.
In Figs.~\ref{fig:lattice_and_GS} (c)–(e), three of the six degenerate ground-state configurations of the model. The remaining three are obtained by exchanging two Potts spin states. The sixfold degeneracy of the ground state gives rise to an emergent $U(1)$ symmetry at the phase transition, rendering the system effectively equivalent to the six-state clock model~\cite{6jpk-yj56}, which is the  signal of a BKT transition.

To characterize out-of-equilibrium defect formation under finite-rate quenches, we employ the energy density $E(t)$ as a primary observable. For this Potts model, the total domain-wall density is linearly proportional to the system energy, meaning both quantities share identical scaling behavior with the quench rate. Since spin configurations evolve dynamically during quenches, both observables are time-dependent.

We define the time-dependent defect density $n(t)$ as
\begin{equation}
\label{eq:n(t)}
n(t)=
\frac{
|J_1|\sum_{\langle i,j\rangle}\delta_{\sigma_i,\sigma_j}
+
|J_2|\sum_{\langle\langle i,j\rangle\rangle}\big(1-\delta_{\sigma_i,\sigma_j}\big)
}{2N\left(|J_1|+|J_2|\right)},
\end{equation}
where topological defects correspond to energetically unfavorable broken bonds: a NN bond contributes to the defect density when neighboring spins are aligned ($\sigma_i=\sigma_j$), whereas a diagonal NNN bond contributes when spins are misaligned ($\sigma_i\neq\sigma_j$). Here $N$ is the total number of lattice sites; each site carries four NN/NNN bonds in total, leading to $2N$ pairwise bonds across the full lattice.
We weight each bond by \(|J_1|\) and \(|J_2|\) to match the energy cost of spin mismatches. Here, spin mismatches refer to opposite orientations of nearest-neighbor spins, or parallel alignment of next-nearest-neighbor spins. This weighting makes \(n(t)\) linearly proportional to \(E(t)\) for all spin configurations. This linear relation validates the energy density as a convenient substitute for defect density in our KZM scaling analysis.
The instantaneous energy density $E(t)$ is given by
\begin{equation}
\label{eq:E(t)}
E(t)=\frac{-J_1\sum_{\langle i,j\rangle}\delta_{\sigma_i,\sigma_j}
-J_2\sum_{\langle\langle i,j\rangle\rangle}
\delta_{\sigma_i,\sigma_j}}{N}.
\end{equation}
% Owing to the linear proportionality between domain-wall density and total energy, we use the computationally convenient energy density $E(t)$ as a direct proxy for the defect density $n(t)$ when testing KZM scaling predictions in the following sections.
\subsection{Observables}
In the following, we denote the total energy of the system by $H$. The observables include the following:
\begin{itemize}
\item Equilibrium energy density \( E^{ \rm eq} \), which can be obtained using the relation~\cite{PhysRevLett.86.2050,PhysRevE.64.056101}:
\begin{align}
E^{\rm eq}
=
\frac{1}{N}
\frac{
\sum_{ H}
 H \,
g( H)\,
e^{- H/T}
}
{
\sum_{ H}
g( H)\,
e^{- H/T}
},
\end{align}
where $g(H)$ denotes the density of states associated with the total energy $H$ for a configurtion $\{\sigma _i\}$.
\item Excess energy density \( \delta E(t)\) is calculated from the difference between non-equilibrium and equilibrium energies, obtained via the relation below,
	\begin{align}
	\delta E(t) \equiv E(t) - E^{\rm eq}[T(t)].
	\end{align}
\item Specific heat $C_V$, which can be evaluated from the equilibrium energy fluctuations through, 
	\begin{align}
		C_V = \frac{ \langle  H^2\rangle
             -\langle H\rangle^2}{N T^2}.
		\label{eq:cv}
		\end{align}
    \end{itemize} 
    \begin{itemize}
		\item Staggered magnetization  \( \ M_{s} \), is defined as the sum of contributions from different sub-lattices and calculated by the following formula~\cite{Salas1998ThreeState}:   
		\begin{align}
        M_{s} =\frac{1}{N} \sqrt{(M_{Ax}-M_{Bx})^2+ (M_{Ay}-M_{By})^2}, 
		\label{eq:chi}
		\end{align}
        where the components $M_{Ax}$ and $M_{Ay}$ are calculated from spin angles on sub-lattice $A$ as  $M_{Ax} = \sum_{j} cos \uptheta_j$ and $ M_{Ay} = \sum_{j} sin \uptheta_j$. The same expressions apply to sub-lattice $B$ with summation restricted to their respective sites.
	\end{itemize}
	\begin{itemize}
		\item Susceptibility  \( \chi_s\), indicates the fluctuation of magnetization
		which is calculated by the following formula:    
		\begin{align}
			\chi_s = \frac{\langle M_{s}^2 \rangle - \langle M_{s} \rangle^2}{ NT}.
			\label{eq:chi}
		\end{align}
%		where the $M$ represent staggered magnetization $M_{s}$.
	\end{itemize}

We obtain equilibrium thermodynamic quantities, including the equilibrium energy $E^{\mathrm{eq}}$ and specific heat $C_V$, via Wang–Landau Monte Carlo (MC) simulations~\cite{PhysRevLett.86.2050,PhysRevE.64.056101} on a 2D square lattice. Most calculations are performed at linear system size $L=256$, while additional simulations at $L=128$ are carried out for finite-size effect analysis. Compared with conventional MC methods, the Wang–Landau algorithm allows efficient evaluation of equilibrium observables over arbitrary temperature ranges once the density of states is accurately determined for a given lattice size. The magnetic susceptibility $\chi$ is computed using standard Metropolis MC simulations. For each temperature point, the system is first thermalized for $10^5$ MCS, followed by $2 \times 10^5$ MCS for data sampling. To mitigate statistical autocorrelation, measurements of the observables are recorded every 50 MCS.

For nonequilibrium quench dynamics, we adopt the standard Metropolis MC algorithm. To guarantee reliable statistical accuracy, all dynamical quantities are averaged over $1000$ independent quench trajectories, each starting from a fully thermalized configuration with a unique random seed. 

For each process, the system is first equilibrated at the initial temperature $T_{\mathrm{init}}$ for $5\times L^2$ MC steps (MCS). Here, one MCS is defined as $L^2$ single-spin update attempts. After initialization, the system is linearly cooled with a fixed total quench time. During the dynamic quenching process, measurements are performed at each intermediate temperature, with one MCS implemented for data sampling and no additional equilibration applied between successive temperature steps.
\section{Results}
\label{sec:results} 
\subsection{Equilibrium Phase Diagram and Quenching Protocols}
\label{subsec:BKT_transition} 
\begin{figure}[!t]
\includegraphics[width=0.45\textwidth]{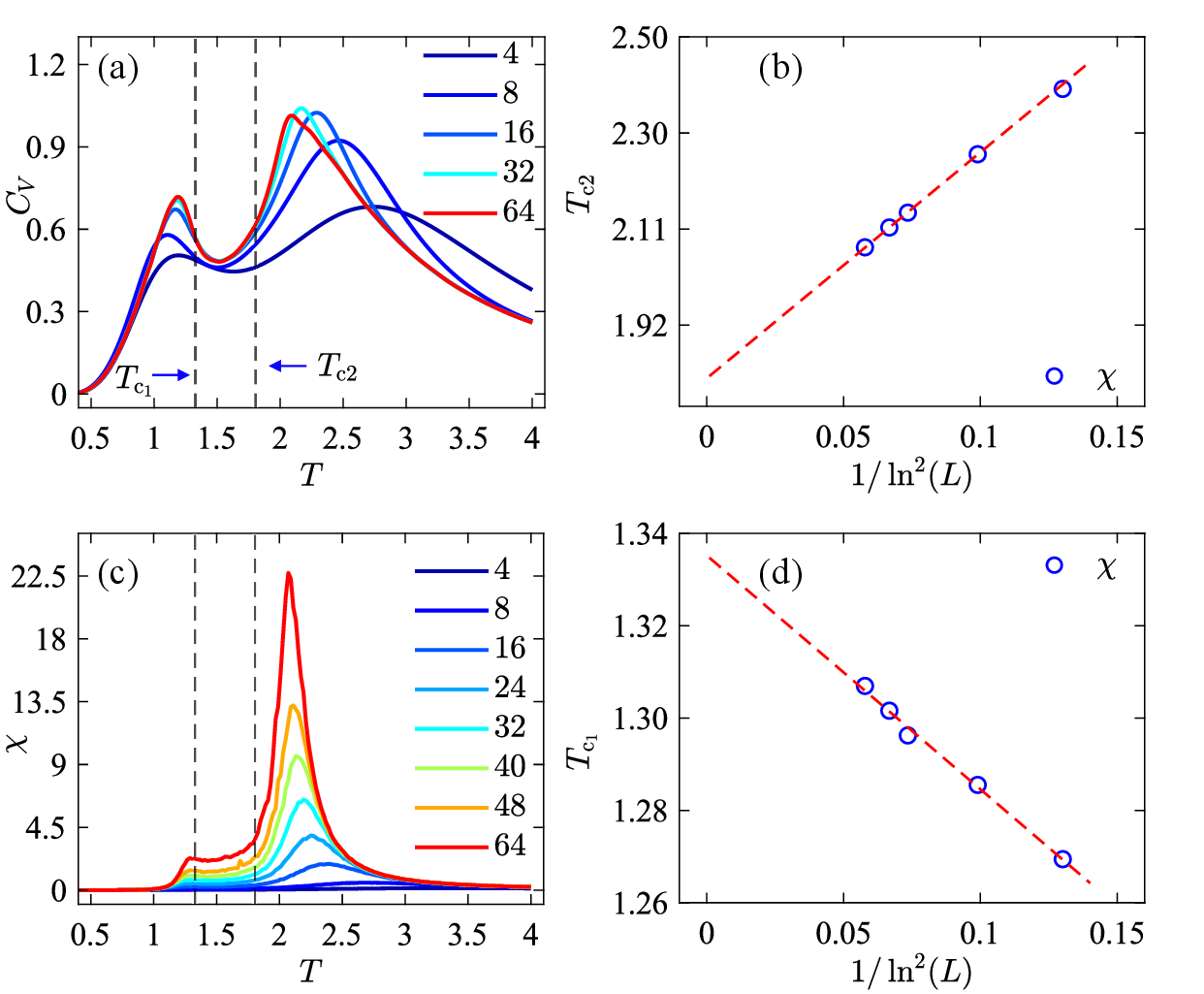}
	\caption{(a) Specific heat $C_V$ versus temperature for different lattice sizes. (c) Magnetic susceptibility $\chi$ versus temperature for different lattice sizes. Vertical dotted lines mark the two thermodynamic-limit BKT critical temperatures. (b) Finite-size scaling of the high-temperature critical point $T_{c2}$ fitted via Eq.~(\ref{T_{c2}_fitting}). (d) Finite-size scaling of the low-temperature critical point $T_{c1}$ fitted via Eq.~(\ref{T_{c1}_fitting}).}
	\label{fig:Total_Cvchi}
\end{figure}

As shown in Figs.~\ref{fig:Total_Cvchi}(a) and (c), consistent with the characteristic feature of BKT transitions, the specific heat does not diverge at critical points and only forms broad, mild peaks. Such broad peak profiles also prevent us from extracting pseudocritical temperatures with high precision. In contrast, susceptibility peaks shift systematically with increasing system size and display prominent finite-size effects. We therefore use susceptibility peak positions to perform finite-size scaling and determine the BKT transition temperatures. The standard BKT finite-size scaling relation, derived from renormalization group theory and holding general validity for all BKT-type topological transitions, is adopted~\cite{2011Computational}:
\begin{equation}\label{eq:T_kt}
T(L) - T_{\mathrm{BKT}} \sim \frac{1}{\ln^{2} L},
\end{equation}
where $T(L)$ denotes the pseudocritical temperature for a lattice of linear size $L$, and $T_{\mathrm{BKT}}$ represents the thermodynamic-limit critical temperature.

We extract size-dependent pseudocritical temperatures from susceptibility data and fit them to the linearized BKT scaling form:
\begin{align}
T_{c1}(L) &= \frac{a_{c1}}{\ln^2 L} + T_{\mathrm{c1,\infty}}, \label{T_{c1}_fitting} \\
T_{c2}(L) &= \frac{a_{c2}}{\ln^2 L} + T_{\mathrm{c2,\infty}}, \label{T_{c2}_fitting}
\end{align}
where $T_{\mathrm{c1,\infty}}$ and $T_{\mathrm{c2,\infty}}$ denote the low- and high-temperature BKT critical temperatures in the thermodynamic limit ($L\to\infty$), respectively. As shown in Figs.~\ref{fig:Total_Cvchi}(b) and (d), the finite-size pseudocritical temperatures exhibit excellent linear scaling with $1/\ln^{2}L$, fully consistent with the canonical BKT scaling behavior. Linear extrapolation to $1/\ln^{2}L \rightarrow 0$ yields $T_{\mathrm{c2,\infty}} = 1.806(1)$ and $T_{\mathrm{c1,\infty}} = 1.329(5)$. 

% The two well-separated peaks in both specific heat and susceptibility confirm the presence of two successive BKT transitions, which naturally divide the system into three temperature-dependent thermodynamic phases. In established double-BKT systems, two consecutive topological transitions always generate an intermediate quasi-long-range ordered (QLRO) phase sandwiched between the high-temperature paramagnetic (PM) phase and the low-temperature long-range ordered (LRO) phase. 
Accordingly, we identify the temperature regime between $T_{\mathrm{c1}}$ and $T_{\mathrm{c2}}$ as the QLRO phase, while the regimes $T>T_{\mathrm{c2}}$ and $T<T_{\mathrm{c1}}$ correspond to PM and LRO phases, respectively.

\begin{figure}[t]
\includegraphics[width=0.45\textwidth]{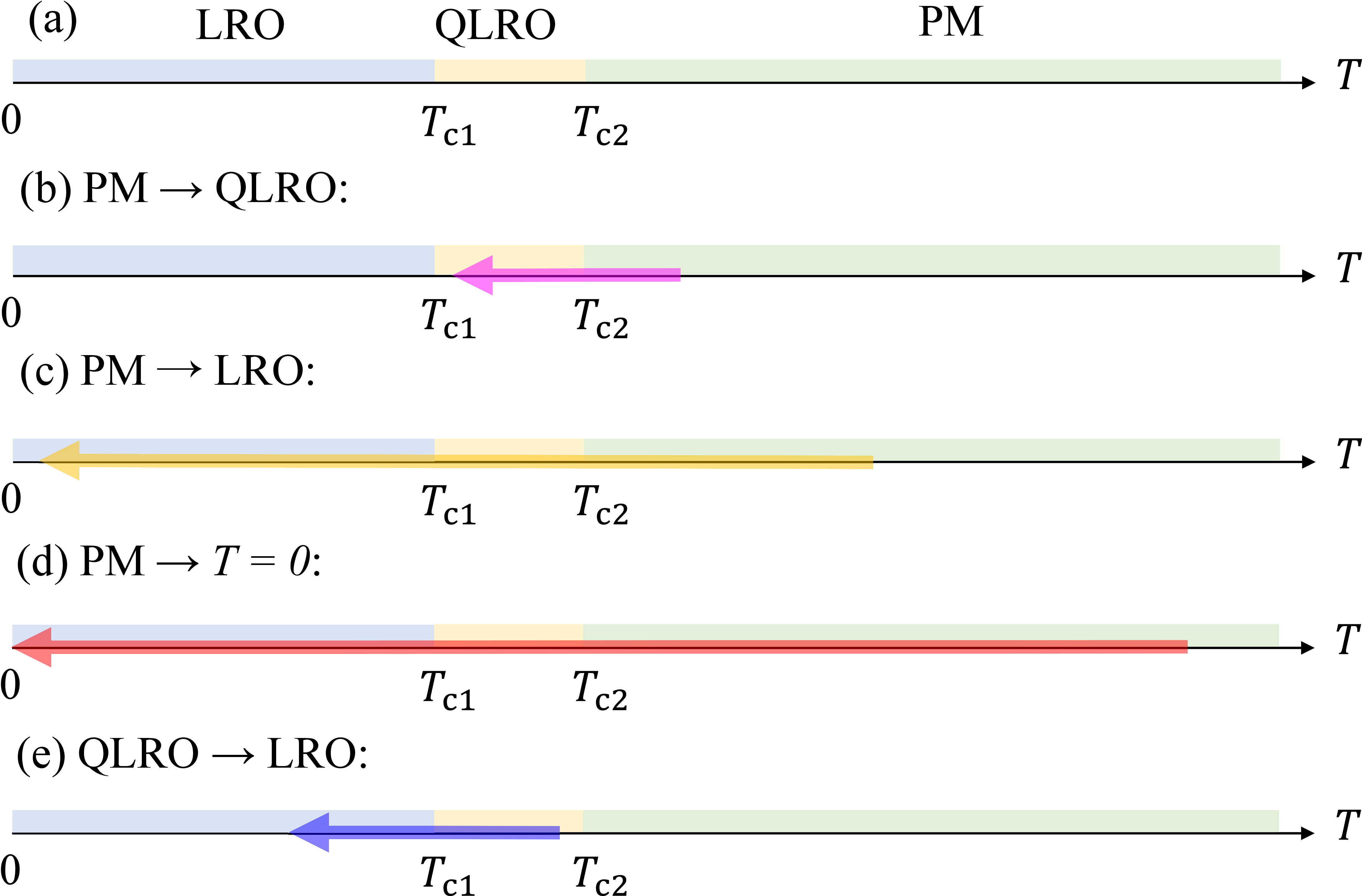}
		\caption{(a) Phase diagram of the $J_1$--$J_2$ Potts model, showing the PM, QLRO, and LRO phases separated by critical temperatures $T_{c1}$ and $T_{c2}$. (b)–(e) Four distinct quenching protocols, where the start and end points of the quenching paths are indicated by pink, yellow, red, and blue arrows, respectively.}
		\label{fig:Three_phase}
	\end{figure}

We take the two critical transition temperatures as symmetric reference points to define the temperature windows for all quench protocols. The four cooling trajectories visualized in Figs.~\ref{fig:Three_phase}(b)–(e) are listed below with their corresponding symmetric reference temperature $T_s$:
\begin{itemize}
\item PM \(\rightarrow\) QLRO:  \(T_{\mathrm{init}}=2.112\),  \(T_{\mathrm{f}}=1.5\), \(T_{\text{s}}=T_{\mathrm{c2}}\);
\item PM \(\rightarrow\) LRO: \(T_{\mathrm{init}}=2.458\), \(T_{\mathrm{f}}=0.2\), \(T_{\text{s}}=T_{\mathrm{c1}}\);
\item PM \(\rightarrow T=0\): \(T_{\mathrm{init}}=3.612\), \(T_{\mathrm{f}}=0\), \(T_{\text{s}}=T_{\mathrm{c2}}\);
\item QLRO \(\rightarrow\) LRO: \(T_{\mathrm{init}}=1.758\), \(T_{\mathrm{f}}=0.9\), \(T_{\text{s}}=T_{\mathrm{c1}}\).
\end{itemize}

For every protocol, the final temperature is set far from both critical boundaries. This arrangement eliminates errors caused by critical fluctuations within the crossover region and guarantees the quench endpoint resides purely inside the target phase.
Three first three  types of finite-rate quench protocols have been extensively explored in existing KZM-related literatures~\cite{reichhardt2022kibble,deutschlander2015kibble,PhysRevB.90.134108,PhysRevB.104.075448,PhysRevE.99.022113,PhysRevE.109.054116,bacsi2023kibble,Schmitt2022Quantum,Jelić_2011}.

\subsection{Quench Dynamics}
\label{sec:QUENCH DYNAMICS}
\subsubsection{\texorpdfstring{PM $\to$ QLRO}{PM → QLRO} quenching protocol}
\label{subsec:PM-QLRO}
We first investigate the quench protocol  in Fig.~\ref{fig:Three_phase}(b), where the system is cooled from the PM phase into the QLRO phase. For the 2D XY model, previous studies have shown that quenches crossing the BKT transition and terminating in the QLRO phase exhibit logarithmic corrections scaling behavior, in contrast to the KZM prediction power-law scaling behavior~\cite{Jelić_2011}. Since the model studied here also possesses an intermediate QLRO phase, a natural question arises: does the PM $\rightarrow$ QLRO quench protocol exhibit the same logarithmic corrections scaling behavior? To address this question, we analyze the quench time $\tau_{Q}$ dependence of the excess energy density $\delta E(t)$ and the nonequilibrium squared staggered magnetization $\langle M_s^2(t)\rangle$, and examine whether these observables follow the logarithmically corrected scaling form.  
%expected for XY-type QLRO dynamics or instead obey the conventional KZM scaling.
\begin{figure}[!tbp]
    \includegraphics[width=0.45\textwidth]{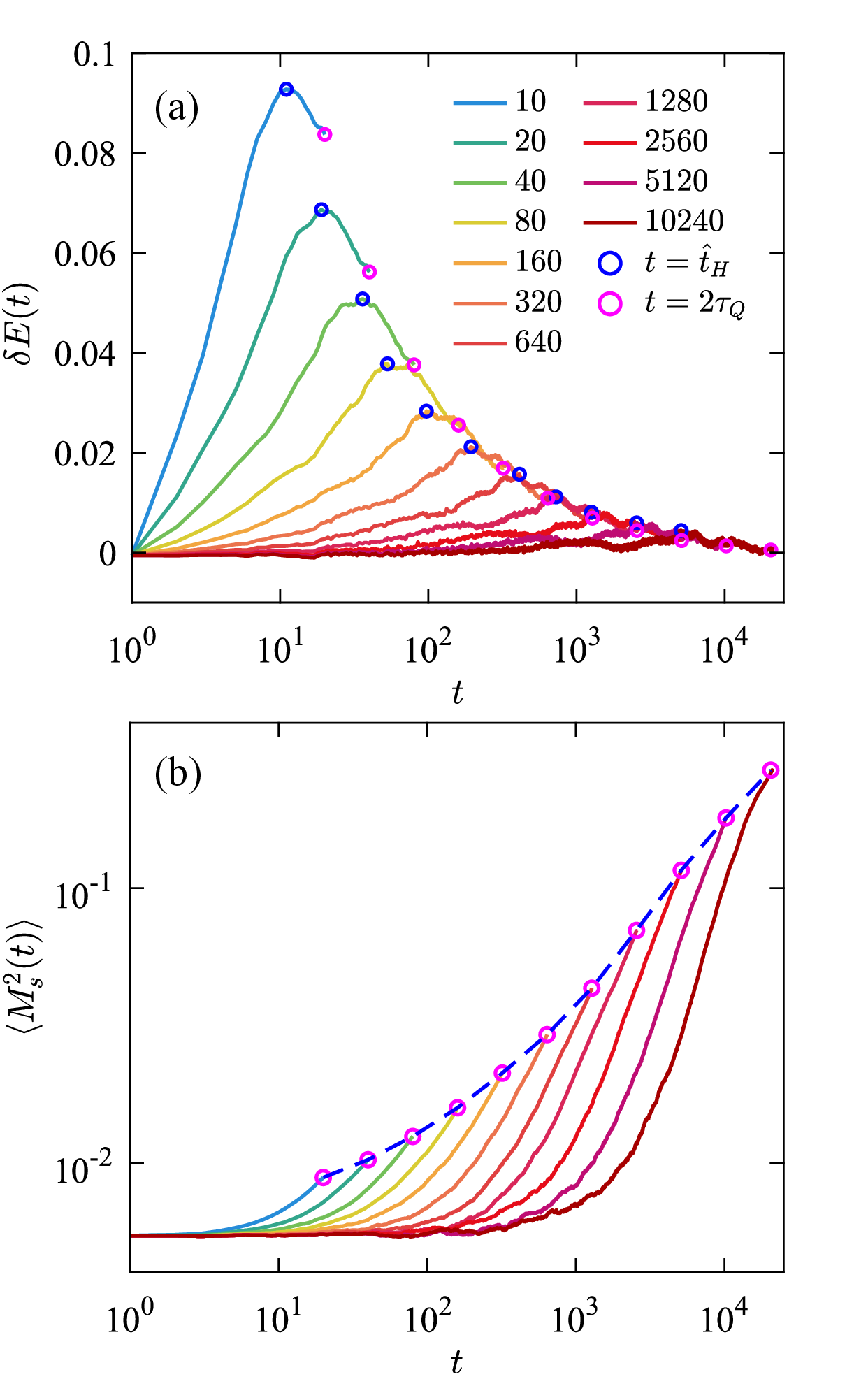}
	\caption{Time evolution of the excess energy density $\delta E(t)$ (panel (a)) and nonequilibrium squared staggered magnetization$\langle M_s^2(t) \rangle$(panel (b)) for different quench times $\tau_Q$. Panel (a) adopts a logarithmic horizontal axis with a linear vertical axis, while panel (b) is plotted on double-logarithmic axes. The blue open circles denote the peak positions at $t=\hat{t}_H$ corresponding to the high-temperature critical point $T_{c2}$ whereas the magenta open circles indicate the termination time $t=2\tau_Q$ of each quench inside the QLRO phase at $T=1.5$.
    }
	\label{fig:T=1_5_quench}
    \end{figure}

As illustrated by the quench process in Fig.~\ref{fig:Three_phase}(b), we employ the linear cooling protocol introduced in Ref~\cite{PhysRevE.99.022113} and choose the temperature range to be symmetric around the critical point $T_{c2}$. The quench protocol is defined as
\begin{equation}\label{eq:quench_protocol}
T(t) = T_{\mathrm{init}} - T_{\mathrm{d}} \, \frac{t}{\tau_{Q}}, \quad t \in [0, 2\tau_{Q}]
\end{equation}
here $T_d$ characterizes the temperature distance between the initial and final temperatures and the critical temperature $T_{c2}$, satisfying
$T_{\mathrm d}\equiv T_{\mathrm{init}}-T_{\mathrm c2}
=T_{\mathrm c2}-T_{\rm f}$. In this section, we set $T_{\mathrm{init}}=2.112$ and $T_{\rm f}=1.5$.

Figure~\ref{fig:T=1_5_quench}(a) shows the temporal evolution of \(\delta E(t)\) for quench times spanning \(\tau_Q=10\) to 10240, covering a wide regime from fast to slow quenches. 
The selected interval of \(\tau_Q\) matches the parameter window adopted in earlier studies of Kibble–Zurek quenching dynamics for the 2D Ising model~\cite{PhysRevE.99.022113} and the Baxter–Wu model~\cite{t1f5-prsf}. For all considered \(\tau_Q\), \(\delta E(t)\) exhibits a universal nonmonotonic behavior, namely increasing first and then decreasing, which forms a prominent peak. This characteristic evolution can be well interpreted based on the definition of the excess energy density, \(\delta E\left(t\right)\equiv E\left(t\right)-E^{\rm eq}[T(t)]\). The entire quench process can be naturally decomposed into three distinct dynamical regimes:\begin{itemize}
\item Adiabatic regime (far from criticality). This regime corresponds to the adiabatic region on the far left of Fig.~\ref{fig:KZ}.
This regime corresponds to the adiabatic region on the far left of Fig.~\ref{fig:KZ}.
At the early quenching stage, the system lies far from the critical region. Here the equilibrium relaxation time $\tau_{\mathrm{eq}}$, plotted as the solid blue line, is far shorter than the external cooling timescale $\tau_t$ (red dashed line) in Fig.~\ref{fig:KZ}. The system can efficiently follow the instantaneous equilibrium state corresponding to the time-dependent temperature, leading to nearly adiabatic evolution. Accordingly, the nonequilibrium energy almost coincides with the equilibrium value, yielding \(\delta E(t)\approx0\).\item  Impulse regime (near criticality). As the temperature approaches the critical region, \(\tau_{\mathrm{eq}}\) grows rapidly and eventually satisfies \(\tau_{\mathrm{eq}}\gg\tau_t\). The system falls out of equilibrium and can no longer adiabatically track the instantaneous thermodynamic state, resulting in significant accumulation of nonequilibrium excitations. The excess energy continuously increases and reaches its maximum around the freeze-out time \(\hat{t}\).
\item Relaxation regime (inside the QLRO phase).
This regime corresponds to the right adiabatic region in Fig.~\ref{fig:KZ}.
After crossing the critical region and entering the ordered QLRO phase, the critical slowing-down effect vanishes, and \(\tau_{\mathrm{eq}}\) becomes much smaller than \(\tau_t\) again. The system gradually relaxes back to the instantaneous equilibrium state, which suppresses the nonequilibrium excitations and causes \(\delta E(t)\) to decay toward zero.\end{itemize}
\begin{figure}[!tbp]
    \includegraphics[width=0.45\textwidth]{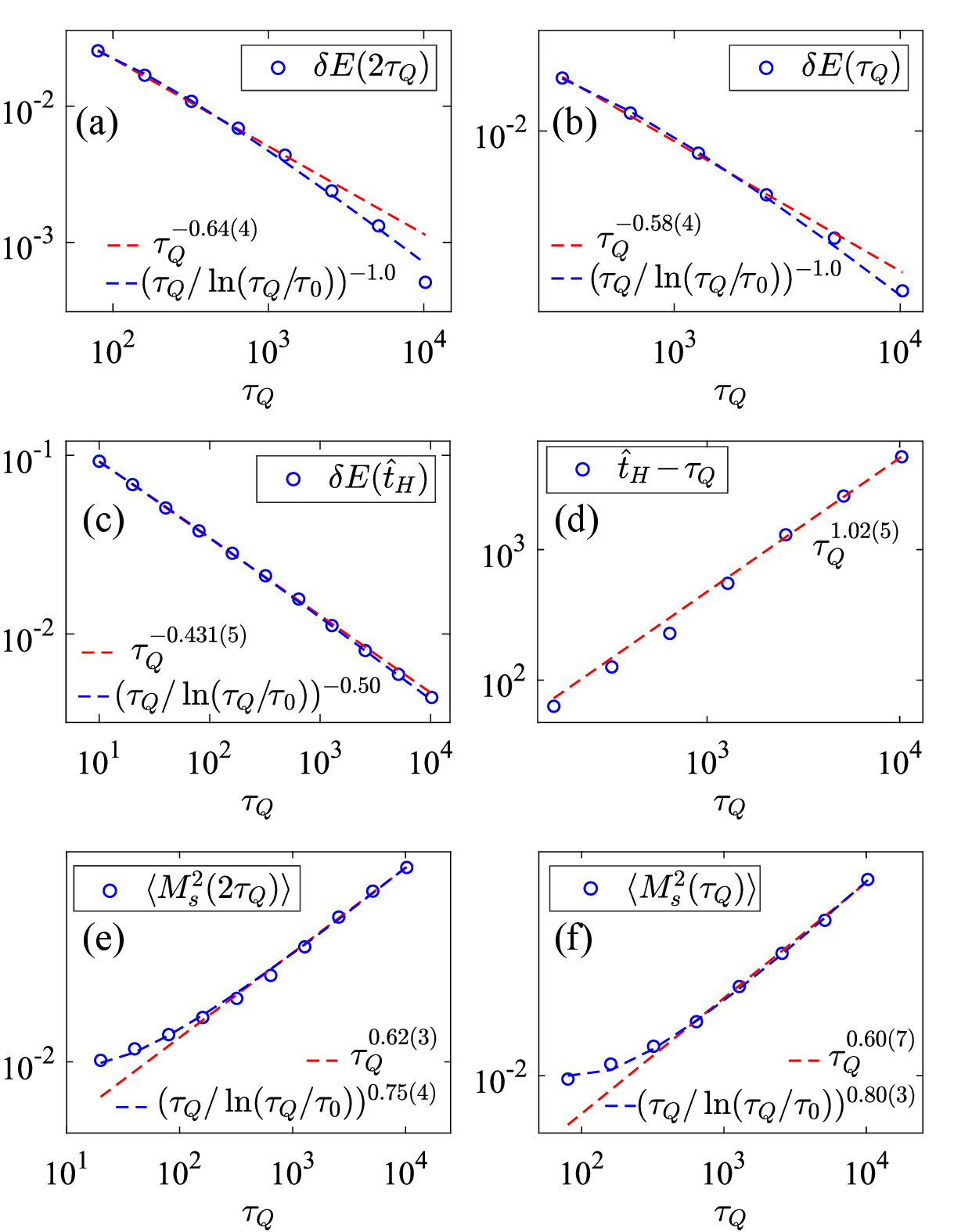}
	% \caption{The scaling of the excess energy density $\delta E(t)$ and the non-equilibrium squared staggered magnetization $\langle M_s^2(t) \rangle$ quenching form $T=2.112$ to $T=1.5$ is presented as follows: (a) $\delta E(t)$ at $t=2\tau_{Q}$ versus $\tau_{Q}$. (b) $\delta E(t)$ at $t=\tau_{Q}$ versus $\tau_{Q}$. (c) $\delta E(t)$ at $t=\hat{t}_H$ versus $\tau_{Q}$. (d) the high-temperature peak positions $t=\hat{t}_H$ versus $\tau_{Q}$. (e) $\langle M_s^2(t) \rangle$ at $t=2\tau_{Q}$ versus $\tau_{Q}$. (f) $\langle M_s^2(t) \rangle$ at $t=\tau_{Q}$ versus $\tau_{Q}$. The blue dashed line represents a scaling with logarithmic correction, while the red dashed line denotes a power-law scaling.}
\caption{Log-log scaling plots for  \(\delta E(t)\) and  $\langle M_s^2(t) \rangle$ under PM-to-QLRO quenches from initial temperature \(T=2.112\) to final temperature \(T=1.5\).
(a) \(\delta E\) at \(t=2\tau_Q\). (b) \(\delta E\) at \(t=\tau_Q\).
(c) $\delta E(t)$ at \(t=\hat{t}_H\). 
(d) the high-temperature peak positions $t=\hat{t}_H-\tau_{Q}$.
(e) $\langle M_s^2(t) \rangle$ evaluated at \(t=2\tau_Q\).
(f) $\langle M_s^2(t) \rangle$ evaluated at \(t=\tau_Q\).
The blue dashed lines correspond to the scaling ansatz with logarithmic correction \((\tau_Q/\ln(\tau_Q/\tau_0))^\alpha\), while the red dashed lines denote simple pure power-law scaling \(\tau_Q^\beta\).}
	\label{fig:T=1_5_fit}
\end{figure}
Moreover, in Fig.~\ref{fig:T=1_5_quench}(a), the peak amplitude of \(\delta E(t)\) declines as \(\tau_Q\) increases. For fast quenches with small \(\tau_Q\), the temperature drops extremely rapidly, so the system cannot follow the instantaneous equilibrium state. A large number of nonequilibrium excitations accumulate, which produces a prominent peak in \(\delta E(t)\). In contrast, slow quenches feature a smaller cooling rate \(dT/dt=-T_d/\tau_Q\) for large \(\tau_Q\). The system gains enough time to relax toward equilibrium during cooling, which suppresses the magnitude of \(\delta E(t)\). Accordingly, the peak height of \(\delta E(t)\) is significantly reduced for larger \(\tau_Q\).

Figure~\ref{fig:T=1_5_quench}(b) plots the curves of the observable $\langle M_s^2(t)\rangle$ versus MC evolution steps for different values of  \(\tau_Q\). At the initial quench time \(t=0\), the system starts from a fully PM phase and is sufficiently relaxed to reach thermal equilibrium. In the thermodynamic limit, the equilibrium value of $\langle M_s^2(t)\rangle$ for the PM phase converges to zero; for the finite lattice size \(L=256\) adopted in this work, the equilibrium $\langle M_s^2(t)\rangle$ at \(t=0\) is approximately $0.005$. As illustrated by all curves, $\langle M_s^2(t)\rangle$ increases monotonically with the number of evolution steps regardless of the chosen \(\tau_Q\). This trend comes from the final QLRO phase that the system relaxes into. At identical temperatures, the equilibrium \(\langle M_s^2(t)\rangle\) of the QLRO phase is inherently larger than the value for the PM phase under finite-size conditions.Moreover, for a fixed range of quench temperatures, a larger \(\tau_Q\) corresponds to a longer relaxation window, allowing the system to evolve closer to its equilibrium state. Consequently, larger values of \(\tau_Q\) yield higher saturated $\langle M_s^2(t)\rangle$.

In the dynamical evolution curves corresponding to different $\tau_Q$ presented above, we select three representative time points to examine whether logarithmic corrections exist in the quench dynamics as shown in Fig.~\ref{fig:T=1_5_fit}: the quench endpoint $t=2\tau_Q$, the midpoint of the quench process $t=\tau_Q$, and the freeze-out time $\hat{t}$. Here, $t=2\tau_Q$ corresponds to the final temperature $T=T_{\rm f}$, marking the moment when the system is deep within the QLRO phase. The point $t=\tau_Q$ corresponds to the system temperature $T=T_{c2}$, which is used to investigate the dynamical behavior when the system crosses the critical point. Meanwhile, $\hat{t}$ denotes the time at which $\delta E(t)$ reaches its maximum, and the corresponding temperature is close to, but not exactly equal to, the critical temperature $T_{c2}$. 

% We perform least-squares fits~\cite{LSF_1,LSF_2} to $\delta E(t)$, $\langle M_s^2(t)\rangle$, and $\hat t$, as listed in Table~\ref{tab:QLRO_fit_exponents}, using the assumed scaling functional forms.
% \begin{align}
% &y = a \left[ \frac{\tau_{Q}}{\ln(\tau_{Q}/\tau_{0})} \right]^{-1}, \label{eq:log_correction_fit_1} \\
% &y = a \left[ \frac{\tau_{Q}}{\ln(\tau_{Q}/\tau_{0})} \right]^{b}, \label{eq:log_correction__fit_2} \\
% &y = a \tau_{Q}^{b}, \label{eq:power_law_fit_3}
% \end{align}
% where in the fitting based on Eq.~(\ref{eq:log_correction_fit_1}), the scaling exponent is fixed at $-1$, with the nonuniversal amplitude $a$ and the microscopic time scale $\tau_0$ treated as free fitting parameters; for Eqs.~(\ref{eq:log_correction__fit_2}) and~(\ref{eq:power_law_fit_3}), both the amplitude $a$ and the scaling exponent $b$ are treated as free parameters, while the microscopic time scale $\tau_0$ in Eq.~(\ref{eq:log_correction__fit_2}) is kept fixed and excluded from the fitting procedure.

We carry out least-squares fitting~\cite{LSF_1,LSF_2} for three quantities: $\delta E(t)$, $\langle M_s^2(t)\rangle$, and freeze-out time $\hat{t}$. All fitting results of scaling exponents are summarized in Table~\ref{tab:QLRO_fit_exponents}, with three candidate scaling functional forms adopted:
\begin{align}
&y = a \left[ \frac{\tau_{Q}}{\ln(\tau_{Q}/\tau_{0})} \right]^{-1}, \label{eq:log_correction_fit_1} \\
&y = a \left[ \frac{\tau_{Q}}{\ln(\tau_{Q}/\tau_{0})} \right]^{\alpha}, \label{eq:log_correction_fit_2} \\
&y = a \tau_{Q}^{\beta}. \label{eq:power_law_fit_3}
\end{align}
For fits using Eq.~(\ref{eq:log_correction_fit_1}), the scaling exponent is fixed to $-1$; only the nonuniversal amplitude $a$ and microscopic reference timescale $\tau_0$ are free fitting parameters. For Eq.~(\ref{eq:log_correction_fit_2}) and the pure power-law form Eq.~(\ref{eq:power_law_fit_3}), both the amplitude $a$ and scaling exponent $b$ are left free to vary during fitting. In Eq.~(\ref{eq:log_correction_fit_2}), the microscopic timescale $\tau_0$ is preassigned to a constant value and not treated as a fitting variable.

Figures.~\ref{fig:T=1_5_fit}(a)--(c) show the scaling of $\delta E(t)$ with $\tau_Q$ evaluated at $t = 2\tau_Q$, $\tau_Q$, and $\hat{t}_H$, respectively. The red lines correspond to the power-law fits based on Eq.~(\ref{eq:power_law_fit_3}), and the blue curves to the logarithmic-correction fits based on Eq.~(\ref{eq:log_correction_fit_1}). Owing to the double-logarithmic scale, the logarithmic-correction curves appear approximately linear in the plot. At large $\tau_Q$, the data points clearly deviate from the power-law lines and are well described by the blue logarithmic-correction curves. Fig.~\ref{fig:T=1_5_fit}(d) displays the scaling of $\hat{t}_H$ with $\tau_Q$, where the vertical axis is plotted as the offset $\hat{t}_H - \tau_Q$~\cite{PhysRevE.99.022113}, representing the characteristic time for the system to reach the energy peak after passing through the critical point; the red line corresponds to the power-law fit based on Eq.~(\ref{eq:power_law_fit_3}). Figs.~\ref{fig:T=1_5_fit}(e) and (f) show $\langle M_s^2(t) \rangle$ at $t = 2\tau_Q$ and $\tau_Q$ as functions of $\tau_Q$, respectively. At small $\tau_Q$, the data points significantly deviate from the power-law lines and are well captured by the blue logarithmic-correction curves corresponding to Eq.~(\ref{eq:log_correction_fit_2}).

The fitting parameters and corresponding scaling exponents obtained from the optimal fits are summarized
in Table~\ref{tab:QLRO_fit_exponents}. Below, we discuss these results in detail and compare them with theoretical predictions.
\begin{table}[tbh]
\centering
\caption{
Fit parameters characterizing the quench-time scaling
of $\delta E(t)$, $\langle M_s^2(t) \rangle$, and $\hat t$
for the PM $\rightarrow$ QLRO protocol.
}
\begin{ruledtabular}
\begin{tabular}{lcccc}
%\hline\hline
Quantity & time $t$ & Scaling form & $a$ & $\tau_{0}$ \\  \hline
$\delta E(t)$ & $2\tau_{Q}$     & $\left[ \frac{\tau_{Q}}{\ln(\tau_{Q}/\tau_{0})} \right]^{-1}$ & 1.06(4) & 11(1) \\[6pt]
$\delta E(t)$ & $\tau_{Q}$     & $\left[ \frac{\tau_{Q}}{\ln(\tau_{Q}/\tau_{0})} \right]^{-1}$  & 3.2(1) & 56(4) \\[6pt]
$\delta E(t)$ & $\hat{t}_{H}$  & $\left[ \frac{\tau_{Q}}{\ln(\tau_{Q}/\tau_{0})} \right]^{-0.5}$ & 0.121(3) & 0.029(6) \\[6pt]
$\hat{t}$  & $\hat{t}_{H}-\tau_{Q}$  & $\tau_{Q}^{1.02(5)}$ & 0.4(2) & \\
$\langle M_s^2(t) \rangle$ & $2\tau_{Q}$     & $\left[ \frac{\tau_{Q}}{\ln(\tau_{Q}/\tau_{0})} \right]^{0.75(1)}$ & 0.00024(5) & 6.0 \\[6pt]
$\langle M_s^2(t) \rangle$ & $\tau_{Q}$     & $\left[ \frac{\tau_{Q}}{\ln(\tau_{Q}/\tau_{0})} \right]^{0.80(3)}$ & 0.0003(2) & 35 \\
%\hline\hline
\end{tabular}
\end{ruledtabular}
\label{tab:QLRO_fit_exponents}
\end{table}

The first two rows of Table~\ref{tab:QLRO_fit_exponents} summarize the scaling of $\delta E(t)$ with $\tau_Q$ at $t=2\tau_Q$ and $t=\tau_Q$, respectively. Both show clear logarithmic corrections, deviating from the pure power-law scaling predicted by the standard KZM. Their scaling forms and exponents agree well with the theoretical prediction in Eq.~(\ref{eq:rho_qlro_8}). The third row shows the scaling of $\delta E(t)$ with $\tau_Q$ at $t=\hat{t}_H$. This also follows a logarithmic correction, but with a different scaling exponent. The fourth row presents the scaling of $\hat{t}_H-\tau_{Q}$ with $\tau_Q$, which follows a pure power-law scaling. The fitted exponent agrees well with the theoretical prediction in Eq.~(\ref{eq:hatt}).
The last two rows  summarize the scaling of $\langle M_s^2(t) \rangle$ with $\tau_Q$ at $t = 2\tau_Q$ and $t = \tau_Q$, respectively. Both show logarithmic corrections, but with nonuniversal scaling exponents. 

The scaling behavior of the staggered magnetization \(\langle M_s^2(t)\rangle\) in quench dynamics has been investigated in prior studies. For systems with conventional second-order transitions, simple power-law scaling is theoretically predicted and numerically verified in models including the Ising and Baxter–Wu models~\cite{PhysRevB.90.134108,PhysRevE.99.022113,t1f5-prsf}. However, few investigations address systems hosting QLRO. To our knowledge, logarithmic corrections to the scaling of \(\langle M_s^2(t)\rangle\) during quenches across BKT transitions have not been previously reported for discrete spin lattice models. In the present work, we numerically observe such logarithmic corrections for cooling trajectories terminating in the QLRO phase of the frustrated \(J_1\)–\(J_2\) Potts model. As further validation, y simulations of the pure 2D XY model are presented in Sec.~\ref{appendix:2DXY_fit}, which reveal identical logarithmic corrections to \(\langle M_s^2(t)\rangle\) and demonstrate that this phenomenon is generic to BKT quenches.

%\textcolor{red}{The power-law scaling of magnetization in quench dynamics has been thoroughly derived theoretically, yet no theoretical investigation has been carried out for its logarithmic correction scaling. This work only presents numerical results.}

The scaling results in the first four rows of Table~\ref{tab:QLRO_fit_exponents} reveal that PM\(\to\)QLRO quenches follow the dynamical scaling of the 2D XY model. These scaling similarities suggest that the intermediate QLRO phase shares a common dynamical universality class with the 2D XY model. Both \(\delta E(t)\) and \(\langle M_s^2(t)\rangle\) display logarithmic-corrected scaling.
\subsubsection{\texorpdfstring{PM $\to$ LRO}{PM → LRO} quenching protocol}
\label{sec:PM->LRO} 
 
Next, we investigate the quench process  in Fig.~\ref{fig:Three_phase}(c), where the system is cooled from the PM phase into the LRO phase. This path crosses two transition temperatures, $T_{c2}$ and $T_{c1}$, in sequence, and terminates in the low-temperature LRO phase. Given the logarithmic corrections observed in the PM $\to$ QLRO quench, we address two questions. We investigate whether the scaling upon crossing \(T_{c1}\) carries logarithmic corrections, and what asymptotic form characterizes the system deep in the LRO phase at the quench endpoint. To answer these questions, we analyze the dependence of $\delta E(t)$ and $\langle M_s^2(t) \rangle$ on the $\tau_Q$.

\begin{figure}[t]
     \includegraphics[width=0.45\textwidth]{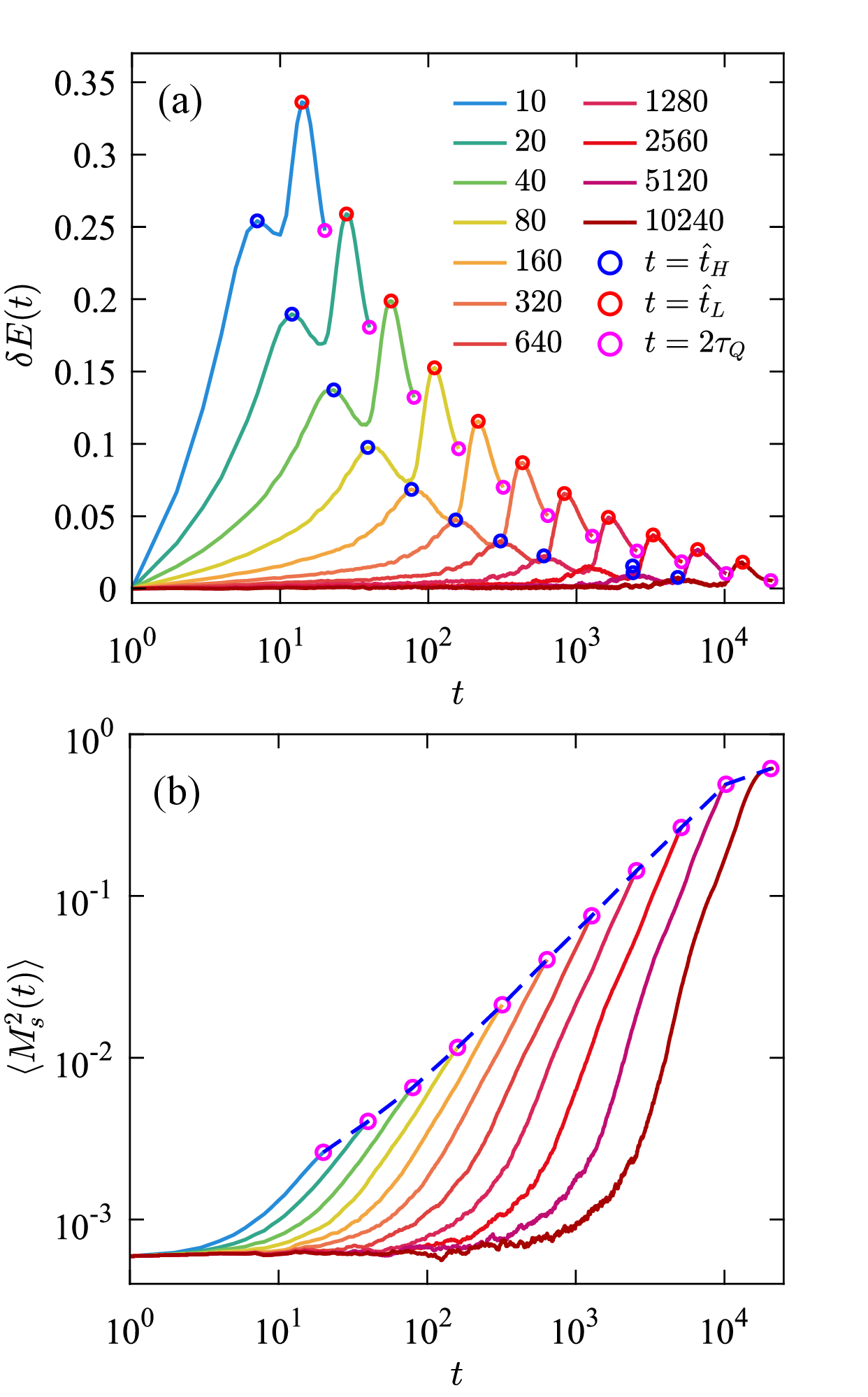 }
	   \caption{Time evolution of  $\delta E(t)$ (panel (a)) and $\langle M_s^2(t) \rangle$ (panel (b)) for different quench times $\tau_Q$. Panel (a) adopts a logarithmic horizontal axis with a linear vertical axis, while panel (b) is plotted on double-logarithmic axes. The blue open circles mark the high-temperature critical peak positions at $t=\hat{t}_H$, the red open circles label the low-temperature critical peak positions at $t=\hat{t}_L$, and the magenta open circles indicate the quenching termination time $t=2\tau_Q$ within the LRO phase at $T=0.2$.}
	    \label{fig:0_2_quench}
  \end{figure}

Following the quench process  in Fig.~\ref{fig:Three_phase}(c), we adopt the same linear quench protocol as in the previous section, except that the temperature interval is set symmetric about the critical point $T_{c1}$, i.e., $T_d \equiv T_{\mathrm{init}} - T_{c1} = T_{c1} - T_{\rm f}$. Here, we choose $T_{\mathrm{init}} = 2.458$ and $T_{\rm f} = 0.2$.
\begin{figure}[!tbp]
     \includegraphics[width=0.45\textwidth]{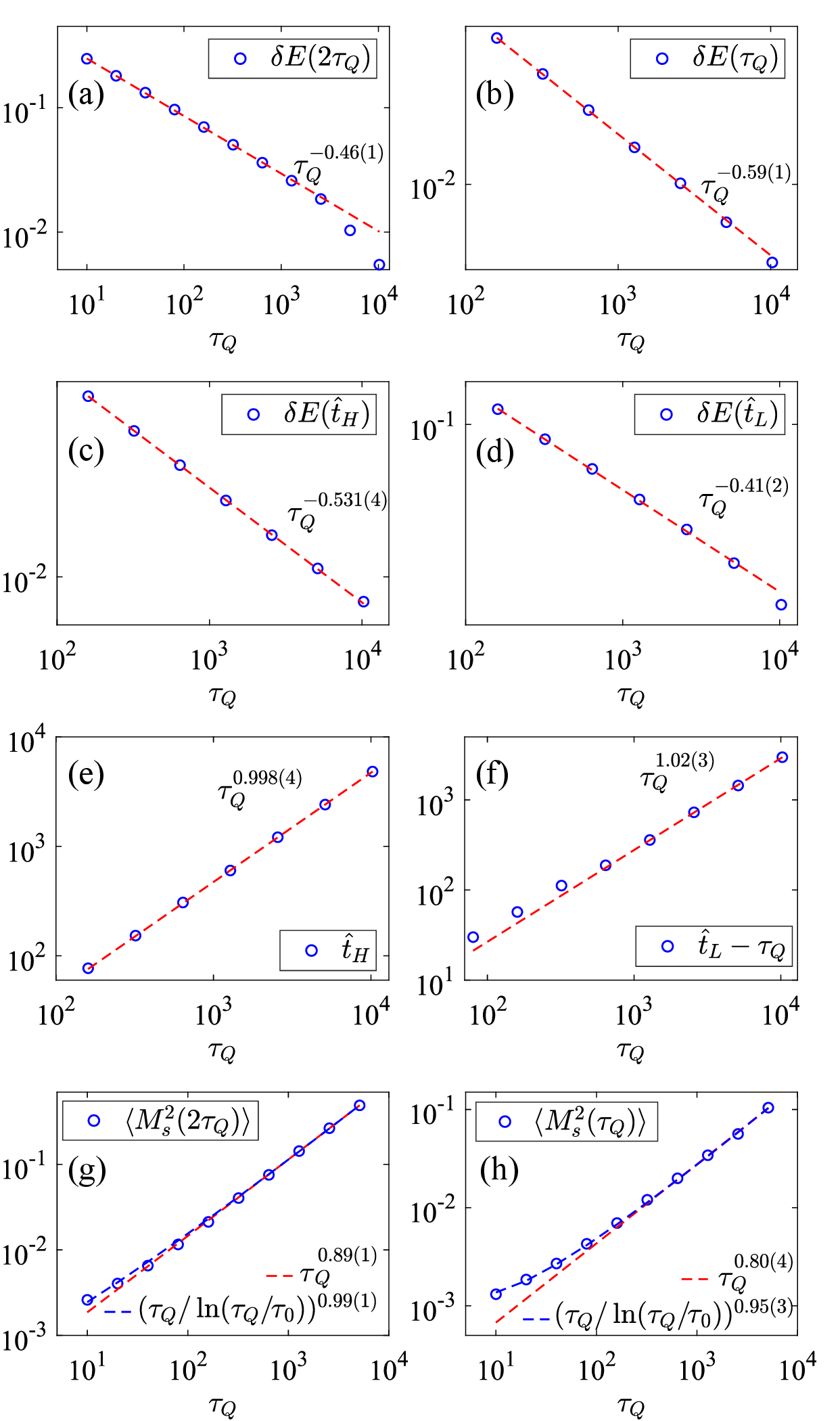}
	   \caption{
       Log-log plots for \(\delta E(t)\) and  \(\langle M_s^2(t)\rangle\) under PM $\to$ QLRO quenches from initial temperature \(T=2.458\) to final temperature \(T=0.2\).
       (a) $\delta E(t)$ at $t=2\tau_{Q}$. (b) $\delta E(t)$ at $t=\tau_{Q}$. (c) $\delta E(t)$ at $t=\hat{t}_H$. (d) $\delta E(t)$ at $t=\hat{t}_L$. (e) the high-temperature peak positions $t=\hat{t}_H$. (f) the low-temperature peak positions $t=\hat{t}_L-\tau_{Q}$. (g) $\langle M_s^2(t) \rangle$ at $t=2\tau_{Q}$. (h) $\langle M_s^2(t) \rangle$ at $t=\tau_{Q}$. The blue dashed line represents a scaling with logarithmic correction, while the red dashed line denotes a power-law scaling.}
	    \label{fig:0_2_fit}
    \end{figure}
    
Figure~\ref{fig:0_2_quench}(a) shows the time evolution of $\delta E(t)$ for different quench times. Unlike in Fig.~\ref{fig:T=1_5_quench}(a), the quench here crosses two transition points, resulting in two distinct peaks. When passing through the lower transition point $T_{c1}$, the system accumulates more nonequilibrium excitations than at the higher transition point $T_{c2}$. As a result, the peak corresponding to $T_{c1}$ is higher than that for $T_{c2}$.

Figure~\ref{fig:0_2_quench}(b) shows curves of $\langle M_s^2(t)\rangle$ against evolution steps for different \(\tau_Q\). The quench path here is PM $\to$ LRO.
At \(t=0\), the system  starts from the PM phase.
For all values of \(\tau_Q\), $\langle M_s^2(t)\rangle$ rises steadily as the number of evolution steps grows. A larger \(\tau_Q\) leads to a longer relaxation process, and the curve reaches a higher maximum value of $\langle M_s^2(t)\rangle$.
This behavior comes from the final LRO phase. The $\langle M_s^2(t)\rangle$ at \(t=2\tau_Q\) in Fig.~\ref{fig:0_2_quench}(b) is larger than $\langle M_s^2(t)\rangle$ for the QLRO final state  in Fig.~\ref{fig:T=1_5_quench}(b).

%\textcolor{red}{Figure~\ref{fig:0_2_quench}(b) presents the evolution of $\langle M_s^2(t)\rangle$. Similar to the PM $\to$ QLRO protocol, the order parameter grows monotonically throughout the quench, while slower quenches result in a larger final magnetization, reflecting the more complete establishment of LRO.}
  
% From the curves at different $\tau_Q$, we select four characteristic times to analyze the scaling behavior of the quench dynamics: the end of the quench $t=2\tau_Q$, the midpoint $t=\tau_Q$, and two freeze-out times $\hat{t}_H$ and $\hat{t}_L$. The times $\hat{t}_H$ and $\hat{t}_L$ correspond to the energy peaks at the higher and lower transition points, respectively. Note that the temperature at $\hat{t}_L$ is very close to, but not exactly equal to, the lower critical temperature $T_{c1}$.
    
\begin{table}[tbh]
\centering
\caption{
Fit parameters characterizing the quench-time scaling
of $\delta E(t)$, $\langle M_s^2(t) \rangle$, and $\hat t$
for the PM $\rightarrow$ LRO protocol.
}
\begin{ruledtabular}
\begin{tabular}{lcccc}
%\hline\hline
Quantity & time $t$ & Scaling form & $a$ & $\tau_{0}$\\ \hline
$\delta E(t)$ & $2\tau_{Q}$     & $\tau_{Q}^{-0.46(1)}$    & 0.72(3) \\[6pt]
$\delta E(t)$ & $\tau_{Q}$      & $\tau_{Q}^{-0.59(1)}$    & 1.09(6) \\[6pt]
$\delta E(t)$ & $\hat{t}_{H}$   & $\tau_{Q}^{-0.531(4)}$   & 1.01(2) \\[6pt]
$\delta E(t)$ & $\hat{t}_{L}$   & $\tau_{Q}^{-0.41(2)}$    & 0.9(1) \\[6pt]
$\hat{t}$     & $\hat{t}_{H}$   & $\tau_{Q}^{0.998(4)}$    & 0.47(1) \\[6pt]
$\hat{t}$     & $\hat{t}_{L}-\tau_{Q}$   & $\tau_{Q}^{1.02(3)}$     & 0.24(6) \\
$\langle M_s^2(t) \rangle$ & $2\tau_{Q}$ & $\left[ \frac{\tau_{Q}}{\ln(\tau_{Q}/\tau_{0})} \right]^{0.99(1)}$ & 0.0009(1) & 0.2\\[6pt]
$\langle M_s^2(t) \rangle$ & $\tau_{Q}$  & $\left[ \frac{\tau_{Q}}{\ln(\tau_{Q}/\tau_{0})} \right]^{0.95(3)}$ & 0.0002(1) & 2.5 \\
%\hline\hline
\end{tabular}
\end{ruledtabular}
\label{tab:fit_exponents_LRO}
\end{table}

Figures~\ref{fig:0_2_fit}(a)--(d) show the scaling of $\delta E(t)$ with $\tau_Q$ at $t=2\tau_Q$, $t=\tau_Q$, $\hat{t}_H$, and $\hat{t}_L$, respectively. The red lines are the power-law fits based on Eq.~(\ref{eq:power_law_fit_3}). In Fig.~\ref{fig:0_2_fit}(a), the data points clearly deviate from the line at large $\tau_Q$, which can be attributed to finite-size effects. In Figs.~\ref{fig:0_2_fit}(b)--(d), the data points agree well with the lines, showing excellent power-law scaling. Figs.~\ref{fig:0_2_fit}(e) and (f) show the scaling of $\hat{t}_H$ and $\hat{t}_L$ with $\tau_Q$, which also follow excellent power-law scaling. In this protocol, the quench crosses two transition points, $T_{c2}$ and $T_{c1}$, with $T_{c1}$ as the center of symmetry. Since $\hat{t}_L$ in Fig.~\ref{fig:0_2_fit}(f) is very close to $\tau_Q$ (the time when the system crosses $T_{c1}$), we plot the offset $\hat{t}_L - \tau_Q$ on the vertical axis~\cite{PhysRevE.99.022113}. Figs.~\ref{fig:0_2_fit}(g) and (h) show the scaling of $\langle M_s^2(t) \rangle$ with $\tau_Q$. The blue curves are the logarithmic correction fits based on Eq.~(\ref{eq:log_correction_fit_2}). At small $\tau_Q$, the data points deviate from the red power-law lines and agree well with the blue logarithmic correction curves.

The fitting parameters and corresponding scaling exponents obtained from the optimal fits are summarized
in Table~\ref{tab:fit_exponents_LRO}. Below, we discuss these results in detail and compare them with theoretical predictions.

The first four rows of Table~\ref{tab:fit_exponents_LRO} show the scaling of $\delta E(t)$ with $\tau_Q$ at $t=2\tau_Q$, $t=\tau_Q$, $\hat{t}_H$, and $\hat{t}_L$, respectively. Since these results follow a pure power-law scaling without logarithmic corrections, we examine whether the fitted exponents agree with the theoretical predictions of the standard KZM. Theoretically, the correlation length critical exponent for the three-state Potts model is $\nu=5/6$~\cite{Baxter:1982zz}, and its dynamical critical exponent is $z=2.171$~\cite{PhysRevE.76.041141,MAydin_1985}.
%\begin{align}z=2.171\left(62\right).\end{align}
Table~\ref{tab:fit_exponents_LRO} lists the numerical results for the power-law exponents. The first row shows the power-law scaling at $t=2\tau_Q$, where the system is deep in the LRO phase. According to Ref~\cite{PhysRevE.99.022113}, the scaling relation deep in the LRO phase is
\begin{align}\delta E\left(2\tau_Q\right)\sim\tau_Q^{-1/z}.\end{align}
Using $z=2.171(62)$, the theoretical prediction is $-1/z \approx -0.46$. This agrees well with the fitted result $-0.46(1)$ in Table~\ref{tab:fit_exponents_LRO}. This indicates that the scaling of $\delta E(2\tau_Q)$ is dominated by the quench dynamics inside the LRO phase at the end of the protocol. Therefore, the observed scaling relation is determined by the dynamical exponent $z$ of the LRO phase, rather than directly reflecting the KZM scaling near the critical point.

%\begin{equation}\label{eq:nu}
 %\nu=5/6  \quad \beta = 1/9 \quad \eta=4/15. 
%  \nu=5/6. 
%\end{equation}
%The dynamics critical exponent $z$ is given by %Refs~\cite{PhysRevE.76.041141,MAydin_1985},
%\begin{equation}\label{eq:Z}
% z = 2.171(62).   
%\end{equation}

%\begin{figure}[!t]  %\includegraphics[width=0.50\textwidth]{M_0_2.eps}
%	   \caption{Nonequilibrium scaling of the squared staggered magnetization $\langle M_s^2(t) \rangle$ versus quench time $\tau_{Q}$ at two characteristic times ($t = 2\tau_{Q}$ and $t = \tau_{Q}$) within the LRO phase ($T_{\rm f} = 0.2$).}
%	    \label{fig:M_0_2}
%\end{figure}
According to the KZM predictions, $\delta E(t)$ and $\hat{t}$ should follow the scaling relations in Eqs.~(\ref{eq:delta ndt_zhengwen}) and~(\ref{eq:thatscaling}), respectively. Based on this, we examine the scaling results in rows two to four of Table~\ref{tab:fit_exponents_LRO}. For $\delta E(t)$, substituting the spatial dimension $D=2$, the defect dimension $d=0$, and the critical exponents $\nu$ and $z$, we obtain the KZM theoretical predictions: $-(D-d)\nu/(1+\nu z) = -0.59$ and $\nu z/(1+\nu z) = 0.64$.

The second row of Table~\ref{tab:fit_exponents_LRO} shows the scaling exponent of $\delta E(t)$ at $t=\tau_Q$. The fitted value is $-0.59(1)$, which agrees well with the KZM prediction of $-0.59$. The third and fourth rows of Table~\ref{tab:fit_exponents_LRO} correspond to the scaling of $\delta E(t)$ at $\hat{t}_H$ and $\hat{t}_L$, respectively. The obtained exponents, $-0.531(4)$ and $-0.41(2)$, both deviate from the theoretical value of $-0.59$.
Standard KZM only describes quenches crossing one critical point, but our cooling passes two BKT transitions. The deviation likely arises because defects created in the first crossing alter the dynamical behavior at \(\hat{t}_H\) and \(\hat{t}_L\).

The fifth and sixth rows of Table~\ref{tab:fit_exponents_LRO} show the scaling of $\hat{t}_H$ and $\hat{t}_L$ with $\tau_Q$. The scaling at $\hat{t}_H$ follows a pure power-law form.  In contrast,  although $\hat{t}_L$ also follows a power-law scaling, its  fitted exponent $1.02(3)$ deviates from the KZM prediction of $0.64$. The seventh and eighth rows of Table~\ref{tab:fit_exponents_LRO} show the scaling of $\langle M_s^2(t) \rangle$ with $\tau_Q$ at $t=2\tau_Q$ and $t=\tau_Q$. At small $\tau_Q$, the data points agree well with the logarithmic-correction curves, showing clear logarithmic corrections.

In summary, $\delta E(t)$ mainly shows power-law scaling, and its exponent at $t=\tau_Q$ agrees well with the KZM prediction. In contrast, $\langle M_s^2(t) \rangle$ still shows logarithmic corrections scaling behavior.
\subsubsection{\texorpdfstring{PM $\to$ $T=0$}{PM to T=0} quenching protocol}
\label{sec:PM_T0}
Next, we consider the quench protocol in Fig.~\ref{fig:Three_phase}(d), where the system is cooled from the PM phase to zero temperature. During this process, it crosses $T_{c2}$, the QLRO phase, and $T_{c1}$, and finally enters the LRO phase. In the LRO phase ($T < T_{c1}$), the system maintains LRO even in the zero-temperature limit. However, it is still unclear whether the dynamical evolution at absolute zero differs significantly from that at any finite temperature ($T \neq 0$). In this section, we address the following question: when the final temperature is set to $T_{\rm f} = 0$, do the scaling laws obtained for quenches to the finite-temperature LRO phase still hold?  To test this hypothesis, we analyze the dependence of the quantities  $\delta E(t)$ and  $\langle M_s^2(t) \rangle$  on the quench time and compare the obtained scaling exponents with the finite-temperature results.
  % \begin{figure}[t]
  %  %  \includegraphics[width=0.5\textwidth]{3_6__0_quench.eps}
  %     \includegraphics[width=0.5\textwidth]{0_0_and_E_M.eps}
	 %   \caption{Time evolution of $\delta E(t)$ (panel (a)) and $\langle M_s^2(t) \rangle$ (panel (b)) for various quench durations $\tau_Q$. Panel (a) is plotted with a logarithmic horizontal axis and linear vertical axis, whereas panel (b) adopts double-logarithmic axes. The blue open circles indicate the high-temperature critical peak positions at $t=\hat{t}_H$, the red open circles label the low-temperature critical peak positions at $t=\hat{t}_L$, and the magenta open circles represent the quenching termination time $t=2\tau_Q$ corresponding to the zero-temperature limit $T=0$.} 
	 %    \label{fig:3_6__0_quench}     
  % \end{figure} 
  \begin{figure}[htbp]
     \includegraphics[width=0.45\textwidth]{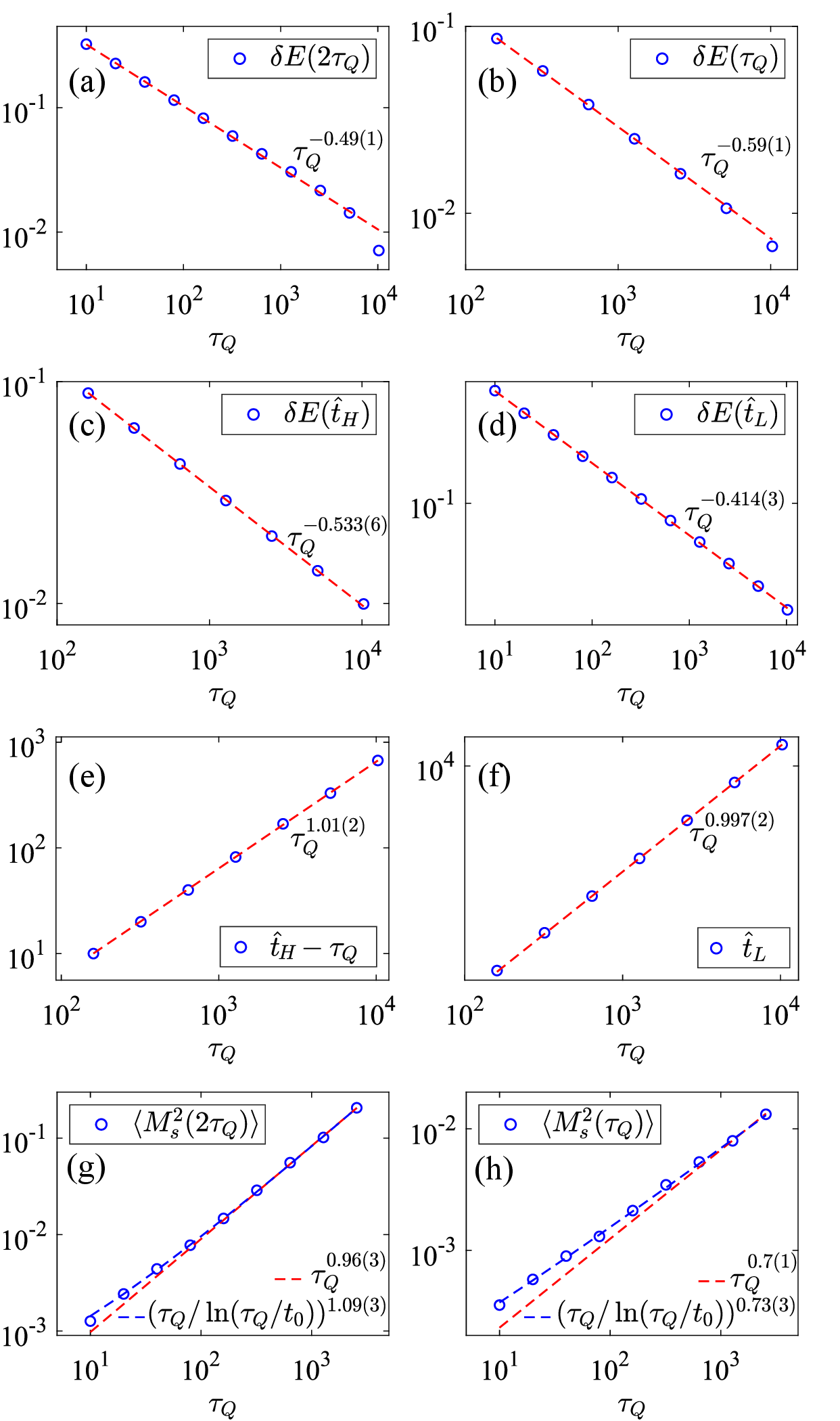}
	   \caption{Log-log plots for  \(\delta E(t)\) and  $\langle M_s^2(t) \rangle$ under PM-to-QLRO quenches from initial temperature \(T=3.612\) to final temperature \(T=0.0\). (a) $\delta E(t)$ at $t=2\tau_{Q}$. (b) $\delta E(t)$ at $t=\tau_{Q}$. (c) $\delta E(t)$ at $t=\hat{t}_H$. (d) $\delta E(t)$ at $t=\hat{t}_L$. (e) the high-temperature peak positions $t=\hat{t}_H-\tau_{Q}$. (f) the low-temperature peak positions $t=\hat{t}_L$. (g) $\langle M_s^2(t) \rangle$ at $t=2\tau_{Q}$. (h) $\langle M_s^2(t) \rangle$ at $t=\tau_{Q}$. The blue dashed line represents a scaling with logarithmic correction, while the red dashed line denotes a power-law scaling.}
	    \label{fig:3_6__0_fit}
  \end{figure}
  
Following the quench process  in Fig.~\ref{fig:Three_phase}(d), we use a linear quench protocol in this section. The temperature interval is set symmetric about the critical point $T_{c2}$, i.e., $T_d \equiv T_{\mathrm{init}} - T_{c2} = T_{c2} - T_{\rm f}$, with $T_{\mathrm{init}} = 3.612$ and $T_{\rm f} = 0.0$.

% % Figure~\ref{fig:3_6__0_quench}(a) shows the evolution of $\delta E(t)$ for different quench times. 
% Similar to Fig.~\ref{fig:0_2_quench}(a), $\delta E(t)$ exhibits two distinct peaks. Because of the temperature range in this protocol is wider and the center of symmetry is different, the system accumulates more nonequilibrium excitations. As a result, the peak values% in Fig.~\ref{fig:3_6__0_quench}(a) 
% are higher than those in Fig.~\ref{fig:0_2_quench}(a). Despite the higher peak values, both quenches show a higher low-temperature peak than the high-temperature one. This indicates that the zero-temperature quench and the finite-temperature LRO quench share similar defect generation processes and nonequilibrium dynamical features. However, we need further scaling analysis to verify whether they belong to the same dynamical universality class.
% % \textcolor{red}{Figure~\ref{fig:3_6__0_quench}(b) displays the evolution of $\langle M_s^2(t)\rangle$. 
% The squared staggered magnetization increases continuously during cooling and reaches larger values for increasing quench time,% indicating that slower quenches allow the system to approach the zero-temperature ordered state more effectively.%}
% This shows that slower cooling lets the system get closer to the fully ordered zero-temperature state.

% From the curves at different $\tau_Q$, we select four times to analyze the scaling of the quench dynamics: the end of the quench $t=2\tau_Q$, the midpoint $t=\tau_Q$, and $\hat{t}_H$ and $\hat{t}_L$.

Figures ~\ref{fig:3_6__0_fit}(a)--(d) show the scaling of $\delta E(t)$ with $\tau_Q$ at $t=2\tau_Q$, $t=\tau_Q$, $\hat{t}_H-\tau_Q$, and $\hat{t}_L$, respectively. The red lines are the power-law fits based on Eq.~(\ref{eq:power_law_fit_3}). In Fig.~\ref{fig:3_6__0_fit}(a), the data points clearly deviate from the line at large $\tau_Q$, which can again be attributed to finite-size effects. In Figs.~\ref{fig:3_6__0_fit}(b)--(d), the data points agree well with the lines, showing excellent power-law scaling. Figs.~\ref{fig:3_6__0_fit}(e) and (f) show the scaling of the freeze-out times with $\tau_Q$, which also follow power-law scaling. Since $\hat{t}_L$ in Fig.~\ref{fig:3_6__0_fit}(e) is very close to $\tau_Q$ (the time when the system crosses the critical point), we plot the offset $\hat{t}_L - \tau_Q$ on the vertical axis. Figs.~\ref{fig:3_6__0_fit}(g) and (h) show the scaling of $\langle M_s^2(t) \rangle$ with $\tau_Q$. The blue curves are the logarithmic-correction fits based on Eq.~(\ref{eq:log_correction_fit_2}). At small $\tau_Q$, the data points deviate from the red power-law lines and agree well with the blue logarithmic-correction curves.

The fitting parameters and corresponding scaling exponents obtained from the optimal fits are summarized
in Table~\ref{tab:fit_exponents_T=0}. Below, we analyze these results in detail.
\begin{table}[tbh]
\centering
\caption{
Fit parameters characterizing the quench-time scaling
of $\delta E(t)$, $\langle M_s^2(t) \rangle$, and $\hat t$
for the PM $\rightarrow$ $T=0$ protocol.
}
\begin{ruledtabular}
\begin{tabular}{lcccc}
% 补全新增的$a$列表头，可按需改为Evaluation time $t$和前表统一
Quantity &time $t$ & Scaling form & $a$  & $\tau_{0}$ \\ \hline
$\delta E(t)$ & $2\tau_{Q}$     & $\tau_{Q}^{-0.49(1)}$    & 1.01(2) \\[6pt]
$\delta E(t)$ & $\tau_{Q}$      & $\tau_{Q}^{-0.59(1)}$    & 1.7(1) \\[6pt]
$\delta E(t)$ & $\hat{t}_{H}$   & $\tau_{Q}^{-0.533(6)}$   & 1.33(4) \\[6pt]
$\delta E(t)$ & $\hat{t}_{L}$   & $\tau_{Q}^{-0.414(3)}$    & 1.15(2)\\[6pt]
$\hat{t}$     & $\hat{t}_{H}-\tau_{Q}$   & $\tau_{Q}^{1.01(2)}$    & 0.059(9) \\[6pt]
$\hat{t}$     & $\hat{t}_{L}$   & $\tau_{Q}^{0.997(2)}$     & 1.47(1) \\
$\langle M_s^2(t) \rangle$ & $2\tau_{Q}$ & $\left[ \frac{\tau_{Q}}{\ln(\tau_{Q}/\tau_{0})} \right]^{1.09(3)}$ & 0.0003(1) & 0.6 \\[6pt]
$\langle M_s^2(t) \rangle$ & $\tau_{Q}$  & $\left[ \frac{\tau_{Q}}{\ln(\tau_{Q}/\tau_{0})} \right]^{0.73(3)}$ & 0.00023(4) & 0.05 \\
%\hline\hline
\end{tabular}
\end{ruledtabular}
\label{tab:fit_exponents_T=0}
\end{table}

The first four rows of Table~\ref{tab:fit_exponents_T=0} show the scaling of $\delta E(t)$ with $\tau_Q$ at $t=2\tau_Q$, $t=\tau_Q$, $\hat{t}_H$, and $\hat{t}_L$, respectively. All the results follow a power-law scaling. Specifically, the exponent in the first row is $-0.49(1)$, which agrees with the value $-0.46(1)$ from the PM $\to$ LRO quench protocol within three error bars. The exponents in the second to fourth rows are $-0.59(1)$, $-0.533(6)$, and $-0.414(3)$, respectively. They agree with the corresponding finite-temperature LRO results of $-0.59(1)$, $-0.531(4)$, and $-0.41(2)$ within one error bar. These results indicate that the zero-temperature and finite-temperature LRO quench protocols share the same scaling behavior.
 
The fifth and sixth rows of Table~~\ref{tab:fit_exponents_T=0} show the scaling of $\hat{t}_H-\tau_{Q}$ and $\hat{t}_L$ with $\tau_Q$. The results follow a power-law scaling. The fitted exponents are $1.01(2)$ and $0.997(1)$, respectively, which agree with the values $0.998(1)$ and $1.02(3)$ from the PM $\to$ LRO quench protocol. The last two rows of Table~\ref{tab:fit_exponents_T=0} show the scaling of $\langle M_s^2(t) \rangle$ with $\tau_Q$ at $t=2\tau_Q$ and $t=\tau_Q$. At small $\tau_Q$, the data points agree well with the logarithmic-correction curves, showing clear logarithmic corrections. This scaling form is the same as that in the PM $\to$ LRO quench protocol.

In summary, although the temperature range, center of symmetry, and peak values in this protocol differ from those in the finite-temperature LRO quench, the scaling behavior and exponents of $\delta E(t)$ are identical. Furthermore, $\langle M_s^2(t) \rangle$ also shows the same logarithmic-correction scaling as in the finite-temperature LRO quench.

\subsubsection{\texorpdfstring{QLRO $\to$ LRO}{QLRO → LRO} quenching protocol}\label{sec:QLRO_to_LRO}
\begin{figure}[t]
      \includegraphics[width=0.45\textwidth]{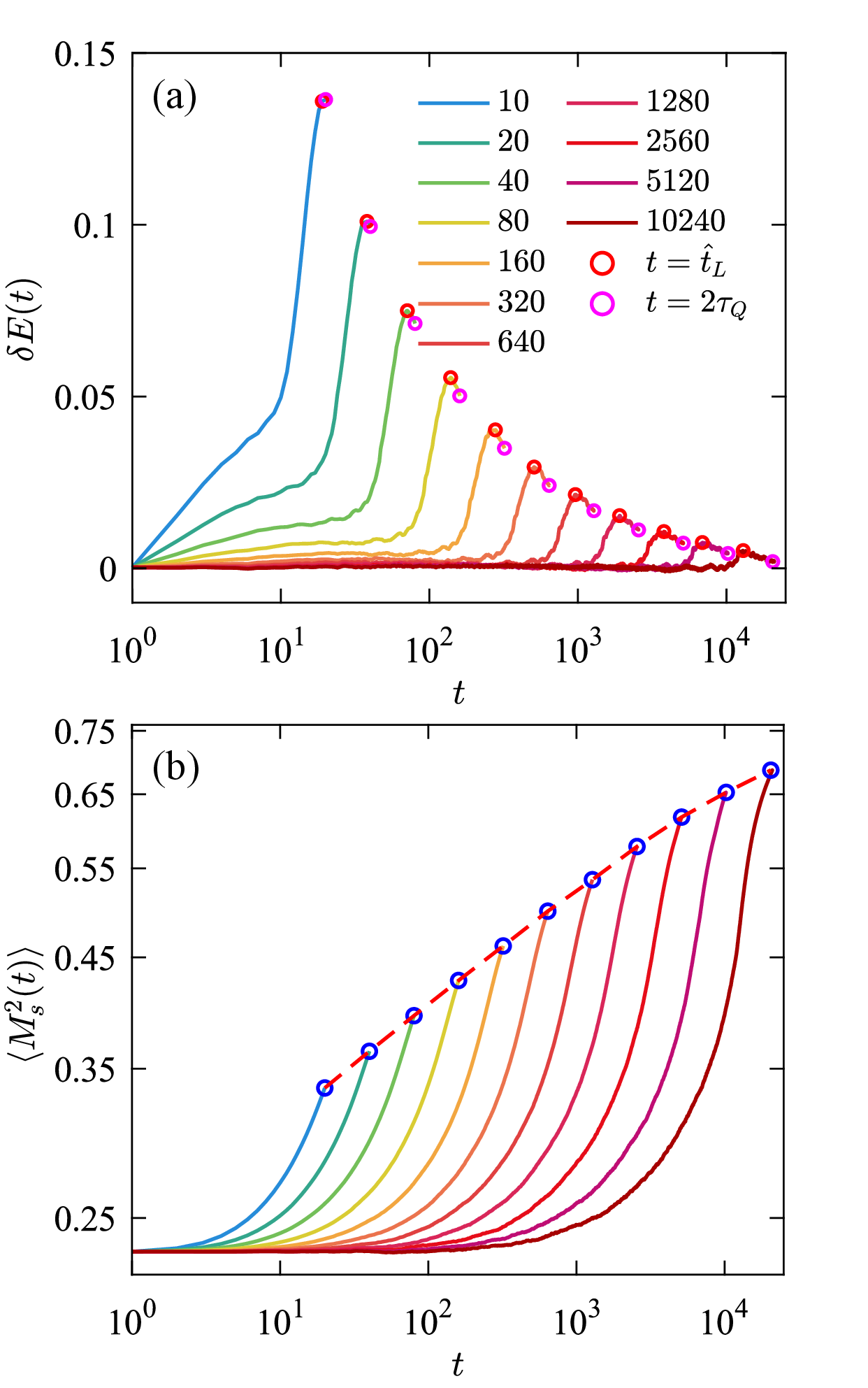}
	   \caption{Time evolution of $\delta E(t)$ (panel (a)) and $\langle M_s^2(t) \rangle$ (panel (b)) for various quench times $\tau_Q$. Panel (a) is plotted with a logarithmic horizontal axis and linear vertical axis, while panel (b) adopts double-logarithmic axes. The red open circles mark the low-temperature critical peak positions at $t=\hat{t}_L$, and the magenta open circles indicate the quenching termination time $t=2\tau_Q$ within the LRO phase at $T=0.9$.}
	    \label{fig:quench_0.9}
    \end{figure}
    
Finally, we study the quench protocol in Fig.~\ref{fig:Three_phase}(e), where the system is cooled from the QLRO phase into the LRO phase. Unlike the previous three protocols, this process avoids the higher transition point $T_{c2}$ and only crosses the lower transition point $T_{c1}$. This allows us to address an important question: is the nonequilibrium scaling mainly related with the initial phase or by the final phase? As in the previous protocols, we analyze the dependence of $\delta E(t)$ and $\langle M_s^2(t) \rangle$ on the quench time.

Following the quench process  in Fig.~\ref{fig:Three_phase}(e), we use a linear quench protocol. The temperature interval is set symmetric about the critical point $T_{c1}$, with $T_{\mathrm{init}} = 1.758$ and $T_{\rm f} = 0.9$.

Figure~\ref{fig:quench_0.9}(a) shows the time evolution of $\delta E(t)$ for different quench times. Unlike the previous PM $\to$ LRO and PM $\to$ $T=0$ protocols, the initial phase here is the QLRO phase. The quench crosses only the lower transition point $T_{c1}$ and ends in the LRO phase. As a result, only one peak appears in the evolution curves. Figure~\ref{fig:quench_0.9}(b) shows the evolution of $\langle M_s^2(t)\rangle$. %Since the system starts from the QLRO phase with finite correlations already established, the order parameter increases smoothly during the quench, and slower cooling leads to a higher degree of LRO in the final state.

% From the curves at different $\tau_Q$, we select three times to analyze the scaling of the quench dynamics: the end of the quench $t=2\tau_Q$, the midpoint $t=\tau_Q$, and $\hat{t}_H$.
 \begin{figure}[t]
    \includegraphics[width=0.45\textwidth]{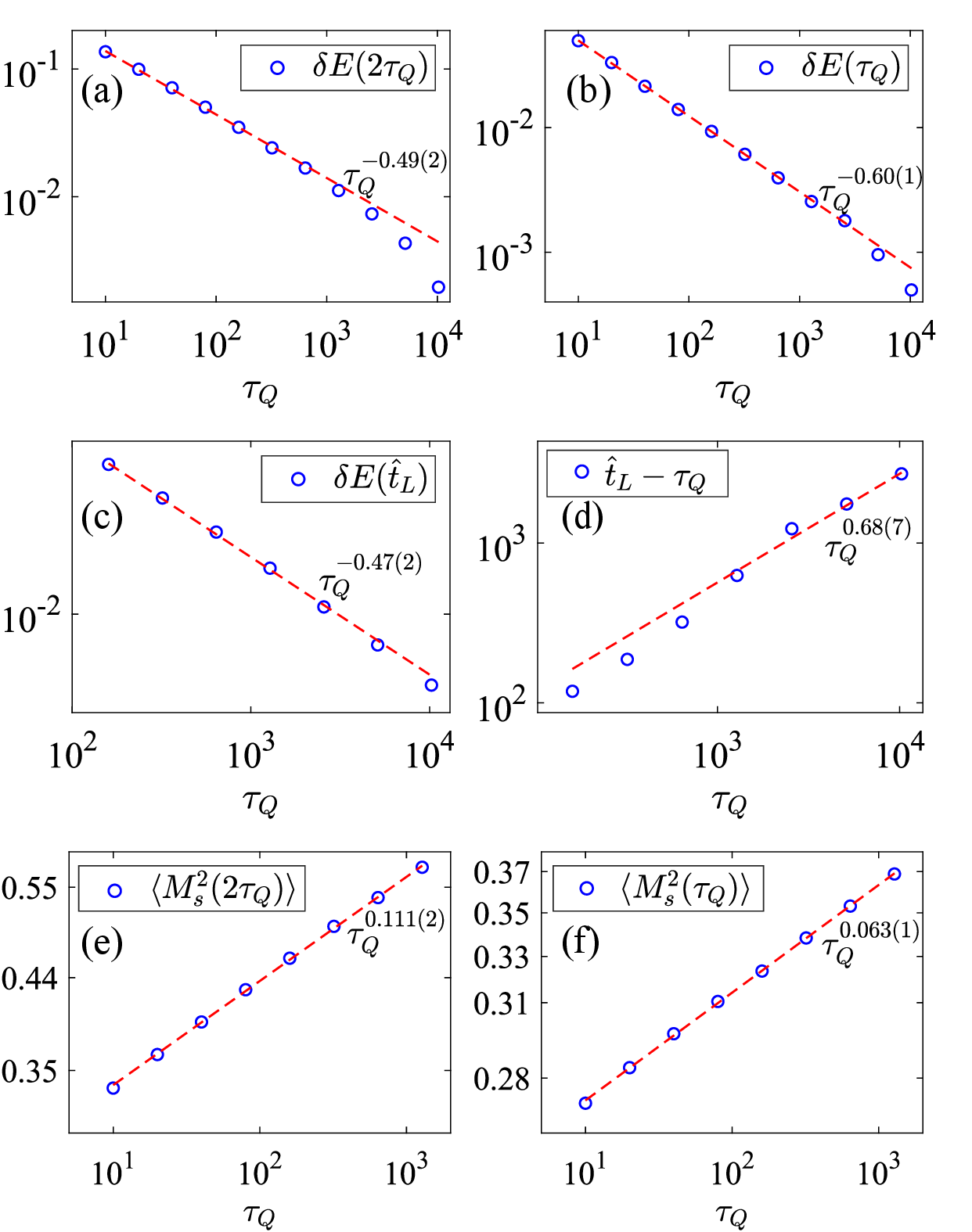}
	   \caption{Log-log  plots for  \(\delta E(t)\) and  $\langle M_s^2(t) \rangle$ under PM-to-QLRO quenches from initial temperature \(T=1.758\) to final temperature \(T=0.9\). (a) $\delta E(t)$ at $t=2\tau_{Q}$. (b) $\delta E(t)$ at $t=\tau_{Q}$. (c) $\delta E(t)$ at $t=\hat{t}_L$. (d) the low-temperature peak positions $t=\hat{t}_L-\tau_{Q}$. (e) $\langle M_s^2(t) \rangle$ at $t=2\tau_{Q}$. (f)  $\langle M_s^2(t) \rangle$ at $t=\tau_{Q}$. The red dashed line denotes a power-law scaling.}
	    \label{fig:0.9_fit}
    \end{figure}  
    
Figures~\ref{fig:0.9_fit}(a)--(c) show the scaling of $\delta E(t)$ with $\tau_Q$ at $t=2\tau_Q$, $t=\tau_Q$, and $t=\hat{t}_L$, respectively. The red lines are the power-law fits based on Eq.~(\ref{eq:power_law_fit_3}). In Fig.~\ref{fig:0.9_fit}(a), the data points deviate from the line at large $\tau_Q$, which can be attributed to finite-size effects. In Figs.~\ref{fig:0.9_fit}(b) and (c), the data points agree well with the lines, showing excellent power-law scaling. Fig.~\ref{fig:0.9_fit}(d) shows the scaling of $\hat{t}_L$ with $\tau_Q$. The data points follow a power-law scaling, and the red line is the corresponding fit. Since $\hat{t}_L$ is very close to $\tau_Q$ (the time when the system crosses the critical point), we plot the offset $\hat{t}_L - \tau_Q$ on the vertical axis. Figs.~\ref{fig:0.9_fit}(e) and (f) show the scaling of $\langle M_s^2(t) \rangle$ with $\tau_Q$ at $t=2\tau_Q$ and $t=\tau_Q$, respectively. Unlike the previous three protocols, the data here show a pure power-law scaling, with the red lines representing the power-law fits.
 
The fitting parameters and corresponding scaling exponents obtained from the optimal fits are summarized
in Table~\ref{tab:fit_exponents_QLRO_LRO}. Below, we analyze these results in detail.
\begin{table}[tbh]
\centering
\caption{Fit parameters characterizing the quench-time scaling
of $\delta E(t)$, $\langle M_s^2(t) \rangle$, and $\hat t$
for the QLRO $\rightarrow$ LRO protocol.
}
% 修正列格式：4列对应lccc（1左对齐+3居中对齐），适配新增的a列
\begin{ruledtabular}
\begin{tabular}{lccc}
% 补全新增的$a$列表头，可按需改为Evaluation time $t$和前表统一
Quantity & time $t$ & Scaling form & $a$ \\ \hline
$\delta E(t)$ & $2\tau_{Q}$     & $\tau_{Q}^{-0.49(1)}$    & 0.43(3) \\[6pt]
$\delta E(t)$ & $\tau_{Q}$      & $\tau_{Q}^{-0.60(1)}$    & 0.20(1) \\[6pt]
$\delta E(t)$ & $\hat{t}_{L}$   & $\tau_{Q}^{-0.47(2)}$    & 0.45(5) \\[6pt]
$\hat{t}$     & $\hat{t}_{L}-\tau_{Q}$   & $\tau_{Q}^{0.68(7)}$     & 5(4) \\
$\langle M_s^2(t) \rangle$ & $2\tau_{Q}$ & $\tau_{Q}^{0.111(2)}$  & 0.26(1) \\ [6pt]
$\langle M_s^2(t) \rangle$ & $\tau_{Q}$ & $\tau_{Q}^{0.063(1)}$  & 0.234(1) \\ [6pt]
%\hline\hline
\end{tabular}
\end{ruledtabular}
\label{tab:fit_exponents_QLRO_LRO}
\end{table}

The first three rows of Table~\ref{tab:fit_exponents_QLRO_LRO} show the scaling of $\delta E(t)$ with $\tau_Q$ at $t=2\tau_Q$, $t=\tau_Q$, and $t=\hat{t}_L$, respectively. All results follow a power-law scaling. The exponent in the first row is $-0.49(1)$, which agrees with the result $-0.46(1)$ from the second quench protocol within three error bars, and with the result $-0.49(1)$ from the third protocol within one error bar. The exponent in the second row is $-0.60(1)$, agreeing with the results $-0.59(1)$ from the second and third protocols within one error bar. The exponent in the third row is $-0.47(2)$, which deviates from the theoretical prediction of $-0.59$.

The fourth row of Table~\ref{tab:fit_exponents_QLRO_LRO} shows the scaling of $\hat{t}_L-\tau_{Q}$ with $\tau_Q$. The result follows a power-law form, and its fitted exponent $0.68(7)$ is close to the theoretical value of $0.64$. The fifth and sixth rows of Table~\ref{tab:fit_exponents_QLRO_LRO} show the scaling of $\langle M_s^2(t) \rangle$ with $\tau_Q$ at $t=2\tau_Q$ and $t=\tau_Q$. They show clear power-law scaling, with the fitted exponents being $0.111(2)$ and $0.063(1)$, respectively.

The results for $\delta E(t)$ agree with those from the second and third protocols. This shows that the scaling behavior and exponents depend only on the final phase, regardless of whether the initial phase is PM or QLRO. In contrast, $\langle M_s^2(t) \rangle$ shows a pure power-law scaling, which is completely different from the first three protocols.
\subsubsection{Finite-size effects and robustness of scaling against parameter choice}\label{sec:Finite-size effect}
 \begin{figure}[t]
     \includegraphics[width=0.45\textwidth]{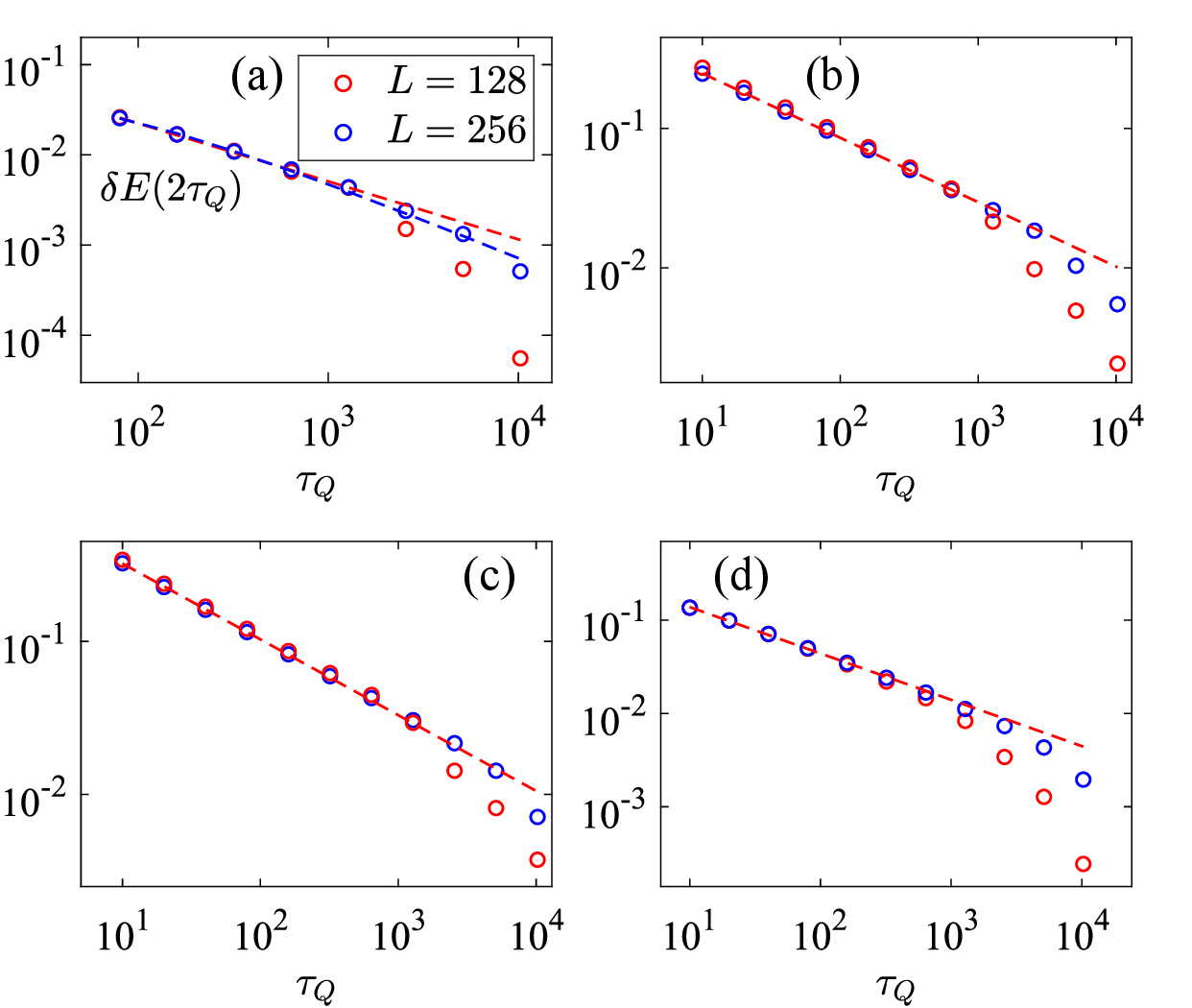}
	   \caption{
Excess energy density $\delta E(t)$  at the quench endpoint $t=2\tau_Q$ for different system sizes.
Panels (a)–(d) correspond to the four distinct quench protocols discussed in the previouse sections.
The red dashed lines represent the power-law fit, and the blue dashed lines represent the logarithmic-correction fit. For larger system sizes, the numerical data points lie closer to the fitting curve.
}
\label{fig:Finite-size effect}
    \end{figure}   
For the above quench protocols, the data at the endpoint $t=2\tau_Q$ deviate from the fitted power-law or logarithmic-correction curves at large $\tau_Q$ as  in Fig.~\ref{fig:T=1_5_fit} (a), Fig.~\ref{fig:0_2_fit} (a), Fig.~\ref{fig:3_6__0_fit} (a) and Fig.~\ref{fig:0.9_fit} (a). 
To show that this is caused by finite-size effects, we compare the results for system sizes $L=128$ and $L=256$ in Fig.~\ref{fig:Finite-size effect}. 

% It can be seen that in Fig.~\ref{fig:Finite-size effect}(a), the data for different sizes overlap at small $\tau_Q$, but the smaller system deviates from the logarithmic-correction fit based on Eq.~(\ref{eq:log_correction_fit_2}) at large $\tau_Q$. 
% Similarly, in Figs.~\ref{fig:Finite-size effect}(b)--(d), the data agree well at small $\tau_Q$ but deviate from the power-law fit based on Eq.~(\ref{eq:power_law_fit_3}) at large $\tau_Q$ for the smaller system. 
%This shows that the deviations are caused by finite-size effects. The systematic drift of the large-$\tau_{Q}$ data with increasing system size suggests that the observed deviations are dominated by finite-size effects.
It can be seen in Fig.~\ref{fig:Finite-size effect}(a) that datasets of different system sizes overlap at small \(\tau_Q\), whereas small-size systems deviate more  from the logarithmic-correction fit of Eq.~(\ref{eq:log_correction_fit_2}) at large \(\tau_Q\). Larger systems produce data points lying much closer to this fitting curve in the large-\(\tau_Q\) region. Similarly, in Figs.~\ref{fig:Finite-size effect}(b)–(d), all datasets collapse onto each other at small \(\tau_Q\), yet small lattices show much larger departures from the power-law fit of Eq.~(\ref{eq:power_law_fit_3}) in the large-\(\tau_Q\) regime. As the lattice size increases, the corresponding data points gradually converge toward the power-law fitting curve.
% \subsubsection{Robustness of the scaling behavior}\label{sec:ROBUSTNESS}

Besides finite-size effects, we also test more cooling processes with different temperature ranges and symmetry centers. Our goal is to confirm that our scaling results are reliable and not accidental matches caused by specific temperature choices.First, we adjust the temperature window for the PM$\to$QLRO cooling process. We set the initial temperature \(T_{\mathrm{init}}=2.212\) and final temperature \(T_{\mathrm{f}}=1.4\).
Second, we adopt a cooling path symmetric around the lower BKT transition \(T_{c1}\) for the PM$\rightarrow$\(T=0\) cooling case. For this setup, the initial temperature is \(T_{\mathrm{init}}=2.658\) and the final temperature is \(T_{\mathrm{f}}=0.0\). In all these extra test conditions, two physical quantities follow the same scaling forms and  exponents shown in the main text: \(\delta E(t)\) and \(\langle M_s^2(t) \rangle\). These observations agree closely with the results presented in previous  sections.
\begin{figure}[t]
    \includegraphics[width=0.45\textwidth]{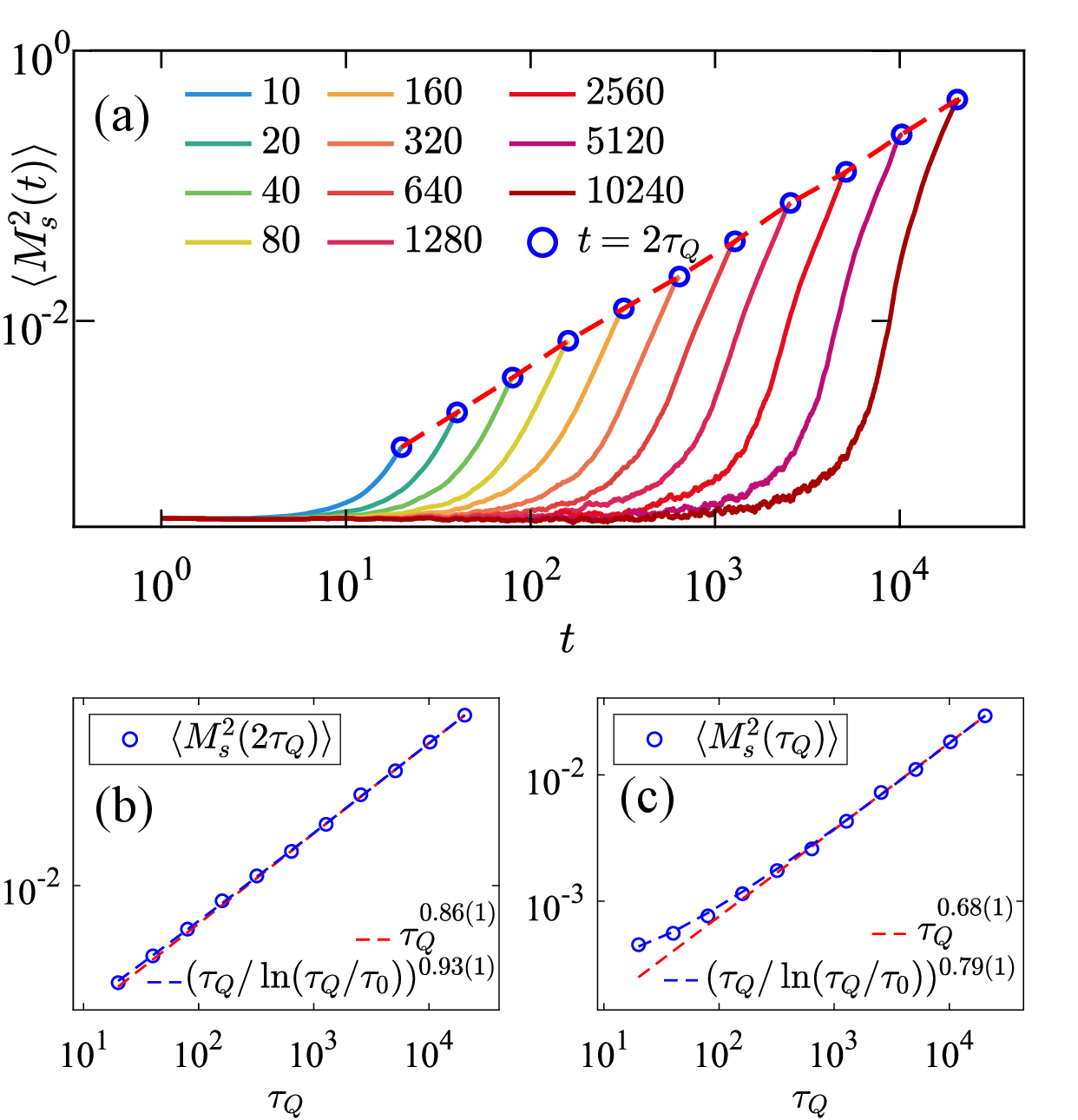}
	\caption{Log-log plots for $\langle M_s^2(t) \rangle$ under PM $\to$ QLRO quenches from initial temperature \(T=1.7858\) to final temperature \(T=0.0\). (a) $\langle M_s^2(t) \rangle$ evaluated at \(t=2\tau_Q\); (b) $\langle M_s^2(t) \rangle$ evaluated at \(t=\tau_Q\).
    The blue dashed lines correspond to the scaling ansatz with logarithmic correction \((\tau_Q/\ln(\tau_Q/\tau_0))^\alpha\), while the red dashed lines denote simple pure power-law scaling \(\tau_Q^\beta\).}
	\label{fig:2DXY_fit}
\end{figure}
\subsubsection{Squared Magnetization in the Nonequilibrium 2D XY Model}\label{appendix:2DXY_fit}

 In the PM $\rightarrow$ QLRO quench protocol of the present work, the nonequilibrium squared magnetization 
  $\langle M_s^2(t)\rangle$ exhibits a logarithmically corrected KZM scaling. %This behavior is naturally associated with the logarithmic correction to the correlation length in the BKT regime, which originates from the logarithmic interaction between vortex-antivortex pairs.
  It is unclear whether this behavior is related to the BKT transition.
We thus investigate if the squared magnetization in the standard XY model shows this feature.
  % Since the two-dimensional XY model is the prototypical system exhibiting a BKT transition and shares the same vortex-driven critical physics, it provides an ideal benchmark for examining the universality of this mechanism. 
  
  % Therefore, in the following Appendix, we investigate the nonequilibrium scaling of the squared magnetization in the PM $\rightarrow$ QLRO quench protocol of the two-dimensional XY model. Observing the same logarithmic correction would provide strong evidence that the scaling behavior identified in the present work is a generic consequence of BKT critical dynamics rather than a model-specific feature of the frustrated Potts model.

 %To further examine the universality of the logarithmically corrected scaling discussed above, we investigate the quench dynamics of the nonequilibrium squared staggered magnetization, $\langle M_s^2(t)\rangle$, in the 2D XY model under the same PM $\rightarrow$ QLRO quench protocol. 
 The complete results are presented in Fig.~\ref{fig:2DXY_fit}. Panel (a) displays the double-logarithmic time evolution of $\langle M_s^2(t)\rangle$ for a wide range of quench times $\tau_Q$, with the blue open circles indicating the measurements at $t=2\tau_Q$. Panels (b) and (c) show the quench-time dependence of $\langle M_s^2(2\tau_Q)\rangle$ and $\langle M_s^2(\tau_Q)\rangle$, respectively. We find that the data at $t=2\tau_Q$ are well described by the logarithmically corrected scaling form $(\tau_Q/\ln(\tau_Q/\tau_0))^{0.93(1)}$, indicating relatively weak logarithmic corrections. In contrast, the data at $t=\tau_Q$ follow the scaling relation $(\tau_Q/\ln(\tau_Q/\tau_0))^{0.79(1)}$, where the logarithmic correction is more pronounced. These results are fully consistent with the behavior observed in the frustrated Potts model.

% This proves our two types of scaling behavior do not change with different cooling paths. One scaling type has logarithmic corrections for the QLRO state, and the other follows a simple power law for the long-range order (LRO) state. Our scaling results stay stable no matter what cooling path we choose.

% To rule out any fortuitous coincidences arising from specific temperature choices and to further validate the robustness of the observed scaling behaviors, we have tested additional quenching protocols with different temperature intervals and symmetry centers. For instance, for the PM $\to$ QLRO, we varied the temperature range to $T_{\mathrm{init}}=2.212$ and $T_{\mathrm{f}}=1.4$. For the PM $\to$ $T=0$ protocol, we employed a cooling path symmetric about the lower BKT transition
% $T_{c1}$ (i.e., $T_{\mathrm{init}}=2.658$ to $T_{\mathrm{f}}=0.0$). In all these alternative setups, the scaling forms and the corresponding asymptotic exponents of both the excess energy density $\delta E(t)$ and the squared staggered magnetization $\langle M_s^2(t) \rangle$ remain highly consistent with those presented in the main text Secs.~\ref{subsec:PM->QLRO} and ~\ref{sec:PM->T=0}. This confirms that the distinct scaling behaviors—whether logarithmically corrected in the QLRO phase or conventional power-law in the LRO phase—are robust and insensitive to the specific microscopic details of the quench trajectories.

\section{Conclusion and Discussion}\label{sec:CONCLUSION}
In this work, we have systematically investigated the equilibrium properties and finite-rate quench dynamics of the 2D three-state $J_1$--$J_2$ antiferromagnetic Potts model by combining equilibrium and nonequilibrium MC simulations. Finite-size scaling of the susceptibility confirms the existence of two successive BKT transitions, which separate the system into the PM, QLRO, and LRO phases.

Our core results reveal that the nonequilibrium scaling behavior of the excess energy density $\delta E(t)$ is  primarily governed by the final phase after quenching,
while the influences stemming from initial spin configurations and detailed quench trajectories appear negligible within our parameter window.  Quenches terminating in the QLRO phase universally exhibit logarithmically corrected KZM scaling, which coincides with the asymptotic scaling behavior of the 2D XY model. By contrast, all quenches ending in the genuine LRO phase, including finite-temperature, zero-temperature, and QLRO$\to$LRO quench routes, follow pure power-law scaling.

The squared staggered magnetization $\langle M_s^2(t) \rangle$ provides complementary information. While its scaling also depends on the quench protocol, logarithmic corrections are observed only for quenches starting from the PM phase, whereas the QLRO$\to$LRO protocol exhibits pure power-law behavior.
Since \(\langle M_s^2(t)\rangle\) equals the spatial integral of spin-spin correlations, we may attribute the logarithmic corrections to long-range spin fluctuations and the distinctive dynamical behavior of the correlation length.

Beyond the specific model studied here, the present framework can be generalized to other frustrated spin systems hosting intermediate quasi-long-range ordered phases.
Future theoretical efforts could extend our nonequilibrium scaling framework to analogous frustrated spin systems, including the six-state clock model, vortex-lattice Potts model, kagome spin ice, and the Ashkin–Teller model~\cite{PhysRevE.80.042103,PhysRevResearch.4.023159,PhysRevB.111.134427,PhysRevLett.116.097206,PhysRevB.108.134422,10.1098/rsta.2011.0388,zhao2026emergentcriticalphasesashkinteller}.
More importantly, rapid advances in programmable photonic platforms have enabled experimental realization of 2D Potts models, offering a promising avenue to directly verify our theoretical and numerical predictions in future experiments~\cite{tgt8-gb13,PhysRevLett.133.237101}.

{\it Acknowledgements.} We are grateful to Prof. C. X. Ding for proposing valuable ideas and for numerous insightful discussions. 
    K. L. and W. Z. were supported by the Hefei National Research Center for Physical Sciences at the Microscale (Grant No. KF2021002) and the Fundamental Research Program of Shanxi Province (Grant Nos. 202303021221029). 
    
\appendix 
\bibliography{references.bib}

\end{document}